# SIMPLE, FAST AND ACCURATE IMPLEMENTATION OF THE DIFFUSION APPROXIMATION ALGORITHM FOR STOCHASTIC ION CHANNELS WITH MULTIPLE STATES

Patricio Orio[1], Daniel Soudry[2,3]

*(1)* Centro Interdisciplinario de Neurociencia de Valparaíso, Facultad de Ciencias, Universidad de Valparaíso, 2360102 Valparaíso, Chile.
*(2)* Department of Electrical Engineering, Technion, Haifa , Israel
*(3)* Laboratory for Network Biology Research, Technion, Haifa , Israel

Address for correspondence:
Dr. Patricio Orio
Centro Interdisciplinario de Neurociencia de Valparaíso
Universidad de Valparaíso
Gran Bretaña 1111
2360102 Valparaiso, Chile
Tel: 56-32-2508185
Fax: 56-32-2508047
e-mail: patricio.orio@uv.cl

# ABSTRACT

**Background**

The phenomena that emerge from the interaction of the stochastic opening and closing of ion channels (channel noise) with the non-linear neural dynamics are essential to our understanding of the operation of the nervous system. The effects that channel noise can have on neural dynamics are generally studied using numerical simulations of stochastic models.

Algorithms based on discrete Markov Chains (MC) seem to be the most reliable and trustworthy, but even optimized algorithms come with a non-negligible computational cost. Diffusion Approximation (DA) methods use Stochastic Differential Equations (SDE) to approximate the behavior of a number of MCs, considerably speeding up simulation times.

However, model comparisons have suggested that DA methods did not lead to the same results as in MC modeling in terms of channel noise statistics and effects on excitability. Recently, it was shown that the difference arose because MCs were modeled with coupled activation subunits, while the DA was modeled using uncoupled activation subunits. Implementations of DA with coupled subunits, in the context of a specific kinetic scheme, yielded similar results to MC. However, it remained unclear how to generalize these implementations to different kinetic schemes, or whether they were faster than MC algorithms. Additionally, a steady state approximation was used for the stochastic terms, which, as we show here, can introduce significant inaccuracies.

**Main Contributions**

We derived the SDE explicitly for any given ion channel kinetic scheme. The resulting generic equations were surprisingly simple and interpretable – allowing an easy, transparent and efficient DA implementation, avoiding unnecessary approximations. The algorithm was tested in a voltage clamp simulation and in two different current clamp simulations, yielding the same results as MC modeling. Also, the simulation efficiency of this DA method demonstrated considerable superiority over MC methods, except when short time steps or low channel numbers were used.

# INTRODUCTION

Noise and variability are present throughout the nervous system, from sensory systems to the motor output and perhaps more importantly in the higher brain areas [1]. Far from being considered as a nuisance, noise is now understood as one of the key elements that shape the way the central nervous system (CNS) codes sensory inputs, builds internal representations and makes decisions [2]. Phenomena like stochastic resonance [3,4,5,6] enhance several aspects of sensory coding and signal detection [7,8]. Also, noise can be beneficial in various computational tasks [9,10,11,12].

One of the main sources of noise and variability is the stochastic opening and closing of ion channels, commonly called *channel noise* [13,14]. The effects of channel noise on neuronal excitability are to a large extent studied with the use of mathematical models, either by constructing and analyzing models with stochastic channels [e.g. 15,16,17,18] or by introducing a noisy conductances in dynamic clamp experiments [19,20]. It is of interest, then, to develop and analyze numerical models that faithfully reproduce the stochastic nature of ion channels. It is also of interest to develop fast algorithms that can be used in large scale simulations of neural networks or in real time simulation for dynamic clamp experiments.

Ion channels are commonly modeled using the framework established by Hodgkin and Huxley [21, see also 22]. In this framework, ion channels contain one or more *activation subunits* that can be either in a resting or active state. The transition rates between states are voltage-dependent, and now we know that this is because these subunits contain a charged domain (the *voltage sensor*) that senses the membrane electrical potential [23]. In the pure Hodgkin and Huxley (HH) framework, the probability of a channel being open is equal to the probability of all its activation subunits being active. Usually the subunits are assumed to be independent and thus the probability of the open channel is the product of the probabilities of the active subunits. In the limit of infinitely many channels (deterministic HH model), probabilities are equivalent to the *fraction* of active subunits or open channels. The transition between resting and active states of subunits is described by ordinary differential equations of a deterministic nature, because the HH model fitted the behavior of a giant squid axon with such a large number of channels that individual stochastic contributions were completely neglected.

When the stochastic behavior of ion channels is taken into account, it is best described by continuous-time, discrete state Markov jumping processes [24,25]. Several algorithms exist for the mathematical simulation of simultaneous and independent Markov Chains (MCs) representing a population of ion channels in a membrane patch or neuronal soma. Among these, the most efficient is a channel-number-tracking algorithm proposed by Gillespie [26] and first applied to ion channels in 1979 [27] (see [28] for a comparison with other MC algorithms). Nevertheless, all MC algorithms increase their computational complexity with the number of channels and the channel-number-tracking algorithm may be difficult to implement for complex kinetic schemes.

Another approach for simulating stochastic ion channels relies on the fact that a large number of simultaneous and independent MCs can be approximated by a stochastic differential equation that describes the time evolution of the fraction of MCs that are in each possible state [29,30,31,32,33]. This algorithm, referred as Diffusion Approximation (DA), is dramatically more efficient in terms of computational cost [28] and is the choice for dynamic clamp experiments where real-time simulation is required [19]. In the general form of DA [29], the time evolution of a variable vector containing the fraction of channels in each state is obtained by solving a Langevin equation (see eq. (1)) with both deterministic and stochastic transition matrices. The method, however, is less practical, since it requires the numerical calculation of a matrix square root at each time step, making it a very time-consuming algorithm (each calculation usually requires about $O(M^3)$ floating point operations [34], $M$ being the number of channel states). To circumvent this, Fox and Lu [29] heuristically proposed to simulate the two-state activation subunits as separate stochastic processes and then calculate the conductance of each ion channel species as the product of subunit probabilities. This approach of uncoupled activation subunits requires a simple Stochastic Differential Equation (SDE) per subunit species without any matrix operation, easily constructed by adding simple noise terms to the deterministic differential equations of the mean channel kinetics. . This, in addition to its high computational efficiency, made the uncoupled subunits approach the main choice for DA implementations [18,19]

However, the uncoupled subunits form of the DA does not approach the behavior of explicit MC appropriately. Mino and colleagues [28] found that this DA algorithm introduces less variability than MC modeling, evidenced as a shallower stimulus vs. action potential firing probability relationship. Later, Bruce [35] found that the DA algorithm, as it was being implemented, assumes that the stochastic term of the gating subunits is uncorrelated, while the MC modeling introduces correlated noise into the channel conductance behavior. Also, the variance of the conductance is higher for MCs than for the uncoupled subunits DA algorithm.

Why was it assumed that activation subunit coupling is of minor importance when modeling stochastic channels? Mainly, because both approaches – coupled or uncoupled subunits – result in the same *mean* time evolution of the conductance. However, fluctuations introduced by both approaches are dramatically different, in terms of the *variance* of the conductances and their *correlations* at different times. This difference between approaches poses a serious problem since the purpose of any quantitative stochastic model is precisely to determine the effects of these fluctuations. The uncoupled subunits approach also has the disadvantage of not being applicable to kinetic schemes with non-independent activation subunits – such as channels with cooperative voltage sensors [36,37], or when the voltage sensors are not identical [38,39].

In recent works [33,40], it was further confirmed that considering coupled activation subunits produces more variability in the conductance and introduces noise with a particular covariance that cannot be reproduced by two-states models. Both works also proposed algorithms for the DA that better approached the results of MC modeling, in the context of the HH model. Goldwyn *et al.* [33] tested the general form of DA suggested by Fox [29], numerically computing the square root of the stochastic diffusion matrix (an $O(M^3)$ operation) at each time step, producing a very time-consuming algorithm. On the other hand, Linaro *et al.* [40] developed a set of SDEs that capture the statistical properties of the variations of conductance, adding it to the ion currents given by a deterministic model.

Here we present a different approach to derive the DA using basic probabilistic tools, for any given kinetic diagram of a channel. This derivations results in a practical, general and intuitive rules allowing for the accurate implementation of DA as a set of simple SDEs, with comparable simplicity to that of (inaccurate) uncoupled DA approach, allowing and efficient implementation (between $O(M)$ and $O(M^2)$ at each time step, depending on the number of kinetic transitions). This makes the computational complexity of the stochastic algorithm comparable to that of the uncoupled DA approach and even the deterministic implementation that simply ignores the noise terms in the SDE. We thoroughly tested the proposed DA implementation, comparing its results to the behavior of explicit MC modeling in three different simulation tests: one under voltage clamp and two under current clamp. Notably, the methods previously suggested [33,40] displayed significant inaccuracies in two of these tests because they employ a steady-state approximation for the calculation of stochastic coefficients. Our method does not require such an approximation and therefore does not incur those errors. We also compare the computational efficiency and numerical stability of the algorithm for different numbers of channels and integration time steps, showing that in most cases DA will be algorithm of choice. Finally, we discuss how our method relates to other implementations previously published.

# RESULTS

We examine a specific population of $N$ ion channels with $M$ states, where the transition rate of a single channel from state $j$ to state $i$ is given by $A_{ij}$. We define the rate matrix $A$ to be composed of these $A_{ij}$ terms for all $i \neq j$, and also $A_{ii} = -\sum_{j \neq i} A_{ji}$ on the diagonal. In neuronal models, these transition rates are usually voltage dependent (and so are also time-dependent). For brevity, we keep this voltage dependency implicit. We denote by $x_i$ the fraction of channels in each of the state, and by $\mathbf{x}$ a vector of $x_i$. Note that $x_1 + \ldots + x_M = 1$ and it is common to use this normalization in order to reduce the number of variables [29,31,33,40]. However, here this substitution is not employed until the numerical implementation to make the algebraic operations easier. The DA proposed by Fox [29,31] for the stochastic dynamics of $\mathbf{x}$ leads to the following SDE

$$\frac{d\mathbf{x}}{dt} = A\mathbf{x} + S\boldsymbol{\xi} ,\qquad(1)$$

where $\boldsymbol{\xi}$ is a vector of independent Gaussian white noise processes with zero mean and unit variance, $A$ is the rate matrix, and $S = \sqrt{D}$, a square root of the diffusion matrix $D$ (namely $SS^\top = D$). This matrix square root has been the main hindrance in the implementation of DA [33]. If solved numerically in simulation time, it incurs a great computational cost, of order $O(M^3)$ at each time step.

Interestingly, it is possible to obtain a direct analytical solution of $S = \sqrt{D}$ for certain kinetic schemes, such as the potassium channel scheme, prior to the simulation (we used Cholesky decomposition, see eq. (15) and below). However, it is not immediately clear how to do so for other schemes, such as the sodium channel scheme. We therefore explored a different derivation of the matrix $S$.

**Derivation of the Diffusion Approximation**

We denote $X_i = Nx_i$, the number of channels in state $i$, and $\mathbf{X}$ to be the corresponding vector. Recall that the channels are independent of each other and that transition rates are memoryless. Therefore, for all $i \neq j$

$$\Delta_{ij}(t) = \left\{\begin{array}{l}\text{the number of channels switching}\\\text{from state } j \text{ to state } i \text{ during } (t, t+dt)\end{array}\right\}\qquad(2)$$

is a Random Variable (RV) composed of the sum of $n = X_j(t)$ independent events ("trials"), in which a channel either switched states, with probability of $p = A_{ij}dt$, or did not switch states, with probability $1 - A_{ij}dt$ (to first order in $dt$). This entails that for all $i \neq j$, $\Delta_{ij}(t)$ are independent and binomially distributed with $n = X_j(t)$ and $p = A_{ij}dt$. Additionally, we define $\Delta_{ii}(t) = 0$. Assume the value of $\mathbf{X}(t)$ is fixed. Denoting by $\langle \cdot \rangle$ the expectation (over the ensemble) we use the properties of the binomial distribution and find the mean

$$\langle \Delta_{ij}(t) \rangle = p = X_j(t) A_{ij} dt , \tag{3}$$

And the variance

$$\mathrm{Var}\left(\Delta_{ij}(t)\right) = np(1-p) = X_j(t) A_{ij} dt \left(1 - A_{ij} dt\right) . \tag{4}$$

Since $\Delta_{ij}(t)$ are independent

$$\mathrm{Cov}\left(\Delta_{ij}(t), \Delta_{mk}(t)\right) = \delta_{im} \delta_{jk} \mathrm{Var}\left(\Delta_{ij}(t)\right) , \tag{5}$$

where $\delta_{ij} = 1$ if $i = j$, and 0 otherwise.

In the limit $N \to \infty, dt \to 0$ we get that $n \to \infty$ and $p \to 0$ for the binomial distribution of each $\Delta_{ij}(t)$. This allows us to approximate $\Delta_{ij}(t)$ by a normal (Gaussian) distribution with both mean and variance equal to $np = X_j A_{ij} dt$ (by the central limit theorem). In order to derive the SDE (eq. (1)), we need to assume that the Gaussian approximation is reasonable. Later, we confirm this numerically, as also did Linaro *et al.* [40] and Goldwyn *et al. [33]* (for example, this was numerically confirmed by [33] for channel numbers as low as $N_K = 18, N_{Na} = 60$).

At each $dt$, $X_i$ changes according to the sum of channels entering and leaving state $i$

$$dX_i(t) = X_i(t + dt) - X_i(t) = \sum_j \left(\Delta_{ij}(t) - \Delta_{ji}(t)\right) . \tag{6}$$

Assuming $\Delta_{ij}(t)$ are all normal, then $d\mathbf{X}(t)$ (the of vector of $dX_i(t)$) is also normal, as a linear combination of independent normal RVs. Since the distribution of normal variables is entirely determined by their mean and covariance, we calculate them.

We use eq. (3) to find the mean of eq. (6)

$$\boldsymbol{\mu}_{d\mathbf{X}}(i) = \langle dX_i(t) \rangle = \sum_j \left(A_{ij} X_j(t) - A_{ji} X_i(t)\right) dt . \tag{7}$$

Next, using eq. (5) we find the covariance

$$R_{dX}(i,j) = \text{Cov}\left(dX_i(t), dX_j(t)\right)$$

$$= \text{Cov}\left(\sum_k \left(\Delta_{ik}(t) - \Delta_{ki}(t)\right), \sum_m \left(\Delta_{jm}(t) - \Delta_{mj}(t)\right)\right)$$

$$= \text{Cov}\left(\sum_k \Delta_{ik}(t), \sum_m \Delta_{jm}(t)\right) + \text{Cov}\left(\sum_k \Delta_{ki}(t), \sum_m \Delta_{mj}(t)\right)$$

$$-\text{Cov}\left(\sum_k \Delta_{ik}(t), \sum_m \Delta_{mj}(t)\right) - \text{Cov}\left(\sum_k \Delta_{ki}(t), \sum_m \Delta_{jm}(t)\right)$$

$$= \delta_{ij}\sum_k \left(\text{Cov}\left(\Delta_{ik}(t), \Delta_{ik}(t)\right) + \text{Cov}\left(\Delta_{ki}(t), \Delta_{ki}(t)\right)\right)$$

$$-\text{Cov}\left(\Delta_{ji}(t), \Delta_{ji}(t)\right) - \text{Cov}\left(\Delta_{ij}(t), \Delta_{ij}(t)\right)$$

$$= \delta_{ij}\sum_k \left(\text{Var}\left(\Delta_{ik}(t)\right) + \text{Var}\left(\Delta_{ki}(t)\right)\right) - \text{Var}\left(\Delta_{ji}(t)\right) - \text{Var}\left(\Delta_{ij}(t)\right)$$

Using the final line of eq. (5), neglecting $dt^2$ terms and dividing by $dt$ we obtain

$$\frac{1}{dt}R_{dX}(i,j) = \begin{cases} \sum_{k \neq i}\left(A_{ik}X_k(t) + A_{ki}X_i(t)\right), & \text{if } i = j \\ -A_{ji}X_i(t) - A_{ij}X_j(t), & \text{if } i \neq j \end{cases}. \tag{8}$$

Since we now know the mean of $d\mathbf{X}(t)$ (eq. (7)) and the covariance between all of its components (eq. (8)), we can write

$$d\mathbf{X} = \mu_{dX} + \sqrt{R_{dX}}\mathbf{Z}, \tag{9}$$

where $\mathbf{Z}$ is a vector of independent Gaussian RVs with mean zero and unit variance. To derive an SDE for $\mathbf{x} = \mathbf{X}/N$ we divide eq. (9) by $N$ and take the limit of $dt \to 0$, yielding

$$\frac{d\mathbf{x}}{dt} = A\mathbf{x} + S\boldsymbol{\xi},$$

which is indeed eq. (1), with $S = \sqrt{D}$, where

$$D_{ij} = \frac{1}{N^2 dt}R_{dX}(i,j) = \frac{1}{N}\begin{cases} \sum_{k \neq i}\left(A_{ik}x_k(t) + A_{ki}x_i(t)\right), & \text{if } i = j \\ -A_{ji}x_i(t) - A_{ij}x_j(t), & \text{if } i \neq j \end{cases}. \tag{10}$$

**A Simpler Derivation of the Diffusion Approximation**

Now that we have the general expression for the diffusion matrix, and know its origin, we can devise a simple way to explicitly calculate $S$, which avoids the use of time consuming numerical procedures for matrix square root computation. The key idea behinds this is to use only $\Delta_{ij}(t)$ and eqs. (3)-(6) to derive the SDE, and the Gaussian approximation. For simplicity, we demonstrate this method step-by-step using a channel with $M = 3$ states

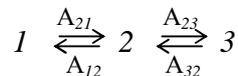

Using eq. (6) we write

$$dX_1 = \Delta_{12} - \Delta_{21}$$
$$dX_2 = \Delta_{21} - \Delta_{12} + \Delta_{23} - \Delta_{32} \quad (11)$$
$$dX_3 = \Delta_{32} - \Delta_{23}$$

Denoting $W_{ij} = \Delta_{ij} - \Delta_{ji}$ we notice that $\Delta_{ij}$ can be combined in opposing pairs

$$dX_1 = W_{12}$$
$$dX_2 = -W_{12} + W_{23} \quad (12)$$
$$dX_3 = -W_{23}$$

We now calculate the means, using $\langle \Delta_{ij}(t) \rangle = X_j(t) A_{ij} dt$ (eq. (3)(7)), we obtain

$$\langle dX_1 \rangle = X_2 A_{12} dt - X_1 A_{21} dt$$
$$\langle dX_2 \rangle = -X_2 A_{12} dt + X_1 A_{21} dt - X_2 A_{32} dt + X_3 A_{23} dt \;.$$
$$\langle dX_3 \rangle = X_2 A_{32} dt - X_3 A_{23} dt$$

Denoting $Y_{ij}(t) = W_{ij}(t) - \langle W_{ij}(t) \rangle$, we obtain

$$dX_1 = \langle dX_1 \rangle + Y_{12}$$
$$dX_2 = \langle dX_2 \rangle - Y_{12} + Y_{23} ,$$
$$dX_3 = \langle dX_3 \rangle - Y_{23}$$

where $Y_{12}, Y_{23}$ are normal, independent, with zero mean and

$$\mathrm{Var}(Y_{12}) = \mathrm{Var}(\Delta_{12}) + \mathrm{Var}(\Delta_{21}) = X_2 A_{12} dt + X_1 A_{21} dt$$
$$\mathrm{Var}(Y_{23}) = \mathrm{Var}(\Delta_{23}) + \mathrm{Var}(\Delta_{32}) = X_3 A_{23} dt + X_2 A_{32} dt ,$$

where we used eq. (5), neglecting $dt^2$ terms. Now we can write

$$dX_1 = X_2 A_{12} dt - X_1 A_{21} dt + Z_1 \sqrt{X_2 A_{12} dt + X_1 A_{21} dt}$$
$$dX_2 = -X_2 A_{12} dt + X_1 A_{21} dt - X_2 A_{32} dt + X_3 A_{23} dt$$
$$\qquad - Z_1 \sqrt{X_2 A_{12} dt + X_1 A_{21} dt} + Z_2 \sqrt{X_2 A_{32} dt + X_3 A_{23} dt}$$
$$dX_3 = X_2 A_{32} dt - X_3 A_{23} dt - Z_2 \sqrt{X_2 A_{32} dt + X_3 A_{23} dt}$$

with $Z_1, Z_2$ are normal, independent, with zero mean and unit variance.

Dividng by $N$ and taking the limit $dt \to 0$, we finally obtain the SDE

$$\frac{dx_1}{dt} = x_2 A_{12} - x_1 A_{21} + \frac{1}{\sqrt{N}} \xi_1 \sqrt{x_2 A_{12} + x_1 A_{21}}$$
$$\frac{dx_2}{dt} = -x_2 A_{12} + x_1 A_{21} - x_2 A_{32} + x_3 A_{23} - \frac{1}{\sqrt{N}} \xi_1 \sqrt{x_2 A_{12} + x_1 A_{21}} + \frac{1}{\sqrt{N}} \xi_2 \sqrt{x_2 A_{32} + x_3 A_{23}}$$
$$\frac{dx_3}{dt} = x_2 A_{32} - x_3 A_{23} - \frac{1}{\sqrt{N}} \xi_2 \sqrt{x_2 A_{32} + x_3 A_{23}}$$

Note that each component of $\xi$ is associated with a transition pair $i \rightleftharpoons j$, multiplied by $\sqrt{(A_{ij}x_j + A_{ji}x_i)/N}$, and appears in the equations of $dx_i/dt$ and $dx_j/dt$ with opposite signs.

Using a similar derivation we can now write $S$ for a general channel with $M$ states. To do this succinctly we must introduce several notations. We denote by $T$ the set of all possible transitions pairs $(i \rightleftharpoons j)$ that exist between states and then give each pair an index in $k = 1,\ldots,|T|$. Note that $|T|$, the size of set $T$, can be any integer between 0 and $M(M-1)/2$. Also, we denote $T(i)$ to be the subset of all transitions pairs that connect to state $i$. Finally, we denote $m_{ik}$ to be the index of the state connected by the $k$-th transition pair, excluding state $i$.

In that case, the matrix $S$ is of size $M \times |T|$, and

$$S_{ik} = \begin{cases} \text{sign}(i - m_{ik}) \dfrac{1}{\sqrt{N}} \sqrt{A_{im_{ik}} x_{m_{ik}} + A_{m_{ik}i} x_i} & , k \in T(i) \\ 0 & , k \notin T(i) \end{cases} \quad (13)$$

**Test Case – Potassium and Sodium Channels**

We have obtained the matrix $S$ analytically, showing that it has a rather simple structure. It is necessary, however, to compare our result with previous definitions of the diffusion matrix as given by Fox [29,31] and used by Goldwyn [33]. For a simple comparison, we will use the case of the potassium channel (see the linear kinetic scheme for coupled subunits in Figure 1). Starting from eq. (1) and defining $\mathbf{x} = \begin{bmatrix} n_0 & n_1 & n_2 & n_3 & n_4 \end{bmatrix}^\top$, the matrix $A_K$ is

$$A_K = \begin{bmatrix} -4\alpha_n & 4\alpha_n & 0 & 0 & 0 \\ \beta_n & -3\alpha_n - \beta_n & 3\alpha_n & 0 & 0 \\ 0 & 2\beta_n & -2\alpha_n - 2\beta_n & 2\alpha_n & 0 \\ 0 & 0 & 3\beta_n & -\alpha_n - 3\beta_n & \alpha_n \\ 0 & 0 & 0 & 4\beta_n & -4\beta_n \end{bmatrix}$$

$S_K$ is defined such that $S_K S_K^T = D$ [29], being

$$D = \frac{1}{N_K} \begin{bmatrix} 4\alpha n_0 + \beta n_1 & -4\alpha n_0 - \beta n_1 & 0 & 0 & 0 \\ -4\alpha n_0 - \beta n_1 & 4\alpha n_0 + \beta n_1 + 3\alpha n_1 + 2\beta n_2 & -3\alpha n_1 - 2\beta n_2 & 0 & 0 \\ 0 & -3\alpha n_1 - 2\beta n_2 & 3\alpha n_1 + 2\beta n_2 + 2\alpha n_2 + 3\beta n_3 & -2\alpha n_2 - 3\beta n_3 & 0 \\ 0 & 0 & -2\alpha n_2 - 3\beta n_3 & 2\alpha n_2 + 3\beta n_3 + \alpha n_3 + 4\beta n_4 & -\alpha n_3 - 4\beta n_4 \\ 0 & 0 & 0 & -\alpha n_3 - 4\beta n_4 & \alpha n_3 + 4\beta n_4 \end{bmatrix}$$

(14)

($n$ sub indices in $\alpha$ and $\beta$ were omitted for abbreviation). Using Cholesky decomposition, we can find $S_K$:

$$S_K = \frac{1}{\sqrt{N_K}} \begin{bmatrix} \sqrt{4\alpha n_0 + \beta n_1} & 0 & 0 & 0 & 0 \\ -\sqrt{4\alpha n_0 + \beta n_1} & \sqrt{3\alpha n_1 + 2\beta n_2} & 0 & 0 & 0 \\ 0 & -\sqrt{3\alpha n_1 + 2\beta n_2} & \sqrt{2\alpha n_2 + 3\beta n_3} & 0 & 0 \\ 0 & 0 & -\sqrt{2\alpha n_2 + 3\beta n_3} & \sqrt{\alpha n_3 + 4\beta n_4} & 0 \\ 0 & 0 & 0 & -\sqrt{\alpha n_3 + 4\beta n_4} & 0 \end{bmatrix}.$$

(15)

Substituting in (1) and performing the matrix operations, the full system of SDE for the $n$ variables can be now written as:

$$\frac{dn_0}{dt} = \left(-4\alpha_n n_0 + \beta_n n_1\right) + \xi_1 \frac{1}{\sqrt{N_K}} \sqrt{4\alpha_n n_0 + \beta_n n_1}$$

$$\frac{dn_1}{dt} = \left(4\alpha_n n_0 - \beta_n n_1 - 3\alpha_n n_1 + 2\beta_n n_2\right) - \xi_1 \frac{1}{\sqrt{N_K}} \sqrt{4\alpha_n n_0 + \beta_n n_1} + \xi_2 \frac{1}{\sqrt{N_K}} \sqrt{3\alpha_n n_1 + 2\beta_n n_2}$$

$$\frac{dn_2}{dt} = \left(3\alpha_n n_1 - 2\beta_n n_2 - 2\alpha_n n_2 + 3\beta_n n_3\right) - \xi_2 \frac{1}{\sqrt{N_K}} \sqrt{3\alpha_n n_1 + 2\beta_n n_2} + \xi_3 \frac{1}{\sqrt{N_K}} \sqrt{2\alpha_n n_2 + 3\beta_n n_3}$$

$$\frac{dn_3}{dt} = \left(2\alpha_n n_2 - 3\beta_n n_3 - \alpha_n n_3 + 4\beta_n n_4\right) - \xi_3 \frac{1}{\sqrt{N_K}} \sqrt{2\alpha_n n_2 + 3\beta_n n_3} + \xi_4 \frac{1}{\sqrt{N_K}} \sqrt{\alpha_n n_3 + 4\beta_n n_4}$$

$$\frac{dn_4}{dt} = \left(\alpha_n n_3 - 4\beta_n n_4\right) - \xi_4 \frac{1}{\sqrt{N_K}} \sqrt{\alpha_n n_3 + 4\beta_n n_4}$$

(16)

where, again, $\xi_1$, $\xi_2$, $\xi_3$ and $\xi_4$ are independent Gaussian white noise terms with zero mean and unit variance. Note in (16) that although the length of the noise vector $\boldsymbol{\xi}$ is equal to the number of states, the number of noise terms actually employed is equal to the number of transition pairs $i \rightleftharpoons j$. Also, as before, each component of $\boldsymbol{\xi}$ is associated with a transition pair $i \rightleftharpoons j$; it is multiplied by $\sqrt{\left(A_{ij} x_j + A_{ji} x_i\right) / N}$, and then added to deterministic differential equations of $dx_i / dt$ and $dx_j / dt$ with opposite signs. Thus, the structure of equations we proposed is also obtained from the original definition of $S$.

However, it is easy to see that Cholesky decomposition, which generates lower triangle matrices, will only work for "linear" kinetic schemes – $1 \rightleftharpoons 2 \rightleftharpoons ... \rightleftharpoons M$. For a example, since a triangle matrix must be square the Cholesky decomposition cannot work if $M < |T|$, as in the case of the sodium channel, where $M = 8$ and $|T| = 10$. In that case, the $S$ matrix we derive is different than that suggested by Fox [29,31] and used by Goldwyn *et al.* [33] – since in the latter approach the length of $\xi$ was always equal to $M$, the number of states and not the number of transition pairs, as in our approach. With our approach, the SDE for sodium channels (see Supplemental Text T1) requires the use of 10 random terms instead of 8 (or 7, if the normalization of $\mathbf{x}$ is used). The use of more stochastic terms may appear computationally more expensive, but it comes with the benefit of simple stochastic equations that avoid complex matrix operations. Finally, it is noteworthy that the $S$ matrix that we propose, with size $M \times |T|$, also fulfills $SS^T = D$, even if $M < |T|$.

In the following sections we will prove that our equations faithfully reproduce the results that can be obtained in simulations with explicit MCs, with similar numerical stability and lower computational cost.

## Numerical Simulations

To test the proposed DA algorithm, it was compared to MC modeling both in their uncoupled and coupled subunits approach. If properly implemented, a DA method considering coupled activation subunits should give the same results as multi-state MCs, while a DA method with uncoupled activation subunits should behave as independent, two-state MCs (see Figure 1). As we show next, this is indeed the case.

Additionally, we examined a common "steady state" approximation employed when using DA methods. In this approximation the variable values in the expressions multiplying the noise terms are replaced by their steady state values [28,31,33,40]. Here we will show that the steady state approximation must be used with great caution depending on the kinetics of the channels simulated.

The details of the specific models we used and the numerical implementation are described in *Methods*. Before we give the simulations results, we clarify a few important numerical issues.

### Numerical implementation issues

An issue that is commonly debated in the implementation of DA is whether to manipulate the state variables to make them increase discretely or to bound them between 0 and 1. Mino *et al.* [28] did both, making the variables to represent an integer number of open channels by multiplying by the number of channels and then rounding them to the lowest integer. Later, Bruce [41] found that rounding to the lowest integer produced a shift of the Firing Efficiency curves to the left, and that it was more appropriate to make the rounding to the nearest integer. In both works the state variables were bounded between 0 and 1 (or between 0 and the number of channels), something that does not impose any mathematical difficulty when dealing with two-state gating subunits.

However, when working with multi-state channels, bounding the variables by manually correcting an off-bound value causes the variable vectors to leave a bounded hyperplane that may cause the diffusion matrix to be no longer positive semi-definite making it impossible to calculate its square root [33]. Therefore, Goldwyn and colleagues decided not to bind the variables and allowed values below 0 and above 1 and instead replaced the variable values in the random terms with their steady state values. We will show here that in some important cases this steady state approximation can introduce significant deviations compared to the exact equations.

In the present work, neither the variables were converted to an integer number of channels nor were they bounded between 0 and 1. The only manipulation performed to ensure real valued random terms was to apply the square root to the absolute value of the argument. As evidenced by the simulations presented here, this did not introduce any noticeable deviation from the simulations with MCs.

**Voltage clamp simulations**

The behavior of the four simulation algorithms was first compared in voltage clamp simulations, using the potassium channel from the HH model alone. The initial condition was the steady state value at -90 mV and a 6 second simulation was performed with the kinetic constants fixed at +70 mV. The number of open channels was recorded at every time step of the simulation (Figure 2A, top, shows 8 simulated traces). 200 independent pulses were simulated and the mean and variance of open channels was calculated for every time step. Figure 2A, middle, shows mean and variance as a function of time and Figure 2A, bottom, shows the relationship between mean and variance of the number of open channels. It is well known that these two moments follow the relationship [42]:

$$\sigma_I^2 = <I> i - \frac{<I>^2}{N} \quad (17)$$

where $\sigma_I^2$ is the variance of the current at any given time, $<I>$ is the mean of the current at the same time, $i$ is the single channel current and $N$ the number of channels. This relationship stems directly from the fact that the current in voltage clamp is the sum of independent binary channel current. In this case, if $p$ is the probability of finding a channel open, then $<I> = Nip$ and $\sigma_I^2 = Ni^2 p(1-p)$, which jointly give eq. (17).

Comparison of Figures 2A and 2B show that the DA, implemented as coupled activation subunits, perfectly reproduces the behavior of MC simulations. In both simulations the fit of the data to Equation (17) yields the correct values of $N$ and $i$. When the voltage clamp simulation is performed with uncoupled subunits models (Fig. 2C and 2D), the variance of the number of open channels increases with a longer delay than the mean, causing the mean vs. variance relationship not to be fit by the inverted parabola. The fit parameters fall very far from the real values regardless of the simulation algorithm. This stems directly from the fact that in the uncoupled subunit approximation the conductance is not the sum of the different channels conductance at a given time, but instead is the multiplication of such sums. Therefore, the derivation of eq. (17) is no longer accurate. However, it is noteworthy that the DA algorithm with uncoupled subunits behaves similar to modeling uncoupled Markov Chains.

The steady state approximation requires the kinetic constants to change slowly compared to the variables. As the kinetic constants are voltage-dependent, the voltage has to change slower than the variables. In a voltage clamp simulation, exactly the opposite happens as the voltage is changed instantaneously at time 0. As expected, simulations that use the steady state approximation performed very poorly, regardless of the activation subunits coupling (Figure 3). In the case of coupled activation subunits (Figure 3A), an almost constant variance of the number of open channels was obtained, and the maximum during the rising phase of the mean was lost. With uncoupled activation subunits (Figure 3B), the maximum in the variance trace was also lost but a longer delay was also observed. As a result, neither model recovered the correct parameters in the mean vs. variance fit.

Thus, our proposed DA algorithm produces the same results as MC modeling. Significant differences appear when subunit coupling is not treated equally in the algorithms, and when steady state approximation is used. We will test it further with current clamp models also assessing the numerical stability and processor time cost.

**Mammalian Ranvier node model**

The performance of the different simulation algorithms in the mammalian Ranvier node (Rb) model [43] was tested using a 1 ms simulation in which a single current pulse of 0.1 ms duration and variable amplitude is given at the beginning (Figure 4A). 1000 simulations are performed at each current amplitude level and the measures of action potential variability (defined in *Methods*, Rb model) are presented in Figures 4B – 4D. There are clearly two pairs of overlapping curves (Rb2MC with Rb2DA and Rb8MC with Rb8DA), indicating that what makes a difference in the behavior of the models is the activation subunit coupling (Rb2 vs. Rb8) and not the numerical algorithm employed (MC vs. DA).

While results in Figure 4 correspond to simulations performed with 1000 channels, simulations were also performed with 500, 5000 and 10000 channels. To present the data in a more concise way, the Firing Efficiency vs. Stimulus amplitude curves were fitted to a cumulative Gaussian distribution (Figure 5A). The mean of the distribution corresponds to the *Threshold*, the stimulus amplitude that has a probability 0.5 of firing an action potential, while the standard deviation ($\sigma$) is a measure of the spread or the input/output relationship. Figure 5B shows the fitting parameters obtained with different number of channels and the tested algorithms. The most relevant observation in these figures is that, like in Figure 3, simulations performed with the same state representation behave the same regardless of the numerical algorithm. In other words, DA reproduces the same behavior that is obtained with MC simulation. Also it is interesting to note that the threshold is almost independent of the number of channels, while $\sigma$ is highly dependent on it. The latter fact is not surprising as fewer channels imply a noisier, more variable simulation and thus a flatter relationship between stimulus amplitude and Firing Efficiency. When more channels are present, noise is reduced and the curve gets steeper, becoming a step function in the deterministic limit (infinite number of channels).

Figure 5C shows a comparison of the DA algorithms with and without the steady state approximation. In the case of uncoupled subunits there is no much difference introduced by this approximation, behaving almost exactly as the exact DA. However, the model with coupled subunits deviates considerably from the exact algorithm, with less variability as evidenced in the lower spread of the activation curves (σ values). Therefore, it seems that the action potential in the Rb model is fast enough to make the steady state approximation not suitable for a model with coupled activation subunits.

*Numerical Stability*

To test and compare the numerical stability of the algorithms presented here, simulations were performed with increased time steps and the effect of time step on the Firing Efficiency curve was observed. Figure 6A shows that as the time step is increased the threshold also increases, indicating a shift to the right of the Firing Efficiency curve. At dt = 10 μs, there is a sudden drop in threshold, but this is probably a sign of a major instability occurring in the numerical integration. An important observation, however, is that all algorithms show the same behavior, reinforcing the idea that our DA algorithm reproduces the behavior of MC modeling. The spread of the Firing Efficiency curve (Figure 6B) remains to a great extent unchanged as dt is increased and once again the simulation algorithm (MC or DA) does not make any difference. In this case, however the state representation makes a difference as the Rb2 model (independent subunits) shows a steeper Firing Efficiency curve than the Rb8 model (coupled subunits). It should be mentioned that when using DA for the coupled subunits approach (Rb8 model) there was a significant number of simulations with dt=5 μs in which an out-of-range voltage value (NaN, ±Inf) was obtained, and all simulations ended out-of-range for dt≥10 μs. This is to some extent avoided if the variables are constrained to be between 0 and 1, but it comes with some computational cost. Normally, this constraint was not imposed in the simulations presented here (nor in the HH model) and for dt≤1 μs it was not necessary at all. Depending on the kinetics of model to be implemented a decision has to be made as to whether it is worth to add a couple of lines of code that will check and correct values out of boundaries.

*Computational cost*

Figure 6C-D plots the time it takes to run 16000 simulations (1000 simulations per stimulus amplitude) in the machine employed for this work, as a function of the integration time step (6C) or the number of channels simulated (6D). It is clear that MC modeling is slower than DA, with all state representations and all conditions tested. On the other hand, the 8-state representation that arises from an independent channel approach is always slower to calculate than its counterpart 2-state representation (independent subunits). This difference is bigger for MC modeling than for the DA algorithm, maybe because this model was tested in an environment mostly oriented to matrix operations. However, the most remarkable observation from Figure 6 is that MC modeling is highly affected by the number of channels in the simulation (more channels imply more transitions to calculate) while the DA method is only sensitive to the time step and completely unaffected by the number of channels.

## Squid axon model

The original Hodgkin and Huxley [21] model for squid giant axon is deterministic and the channel activation functions are continuous variables. In the absence of a stimulus, no action potential is elicited and the system relaxes to a resting voltage very close to -65 mV. However, if discrete stochastic channels are considered spontaneous action potentials arise due to sodium channels fluctuations [16]. Here, two types of stochastic HH models were simulated and the resulting spike frequency and intervals were analyzed. The HH2 model uses the independent subunits approach (2-state activation subunits), while HH58 model uses the coupled subunits approach, with 5-state potassium channels and 8-state sodium channels.

As expected, the frequency of the spontaneous action potentials increases as the number of channels is decreased in all models and simulation algorithms (Figure 7). However, there is a striking difference between the behavior of the HH2 models and the behavior of the HH58 models at the same number of channels. While at $N_{Na}$=1500 the HH2 models barely fires an action potential, the HH58 models fires about 30 action potentials per second. Importantly, our DA algorithm produces the same firing rates as the corresponding MC models. Figure 8A plots the mean action potential frequency observed in the 500 s simulation, as a function of the number of sodium channels ($N_{Na}$) simulated (the number of potassium channels was always set to $N_{Na}$*0.3). The pattern observed with the Rb model is repeated: models with different subunit coupling have a different behavior while the simulation algorithm makes no difference in the results. In order to go beyond the simple firing rate quantification, the Inter-Spike Intervals (ISIs) obtained in each case were plotted in histograms and fitted to an exponential decay function (Figure 8B, also see Eq. (22) in *Methods*). For all ISIs obtained, it was observed that the first two bins (marked with * in the histogram) did not follow the exponential trend so they were excluded when fitting the histograms. This was observed in all simulations and thus it is not caused by a specific simulation algorithm or subunit coupling. Indeed, it has been observed before [18] and is probably due to the resonant properties of the HH model [21,44,45] that, with a frequency of peak response of 67 Hz, will increase the probability of ISIs around 33 ms. Figures 8C and 8D show the fit parameters obtained as a function of the number of sodium channels, and it is evident that the simulation algorithm employed does not make any difference in the ISI distributions, while the subunit coupling does.

As with the Rb model, a DA approximation algorithm was tested in which the variable values of the random term were replaced by their steady-state values. The results obtained with the coupled subunits model (HH58) is plotted in Figure 8 as well (gray triangles). Here the deviations from the exact DA (and MC as well) are minor, probably because the voltage dynamic in this model is slow enough to let the variables (at least the *m* variable) to be at its steady state value during almost all the simulation.

*Numerical Stability*

To check for numerical stability of the methods, simulations were repeated with increasing values of dt, the integration time step. As shown in Figure 9, increasing dt up to 100 μs has little or no effect in the mean rate of spikes (9A) or the parameters of the ISI distribution (9B and 9C). There are some deviations for dt > 10μs, but they are minor compared to what was observed with the Rb model. In this case, no out-of-range voltage values were produced throughout the 500 seconds simulated. Remarkably, the choice of the algorithm has no effect on the numerical stability within the dt values tested.

*Computational cost*

Figure 9D-E plots the time it took to simulate 500 seconds as a function of the time step (9D) and the number of sodium channels (9E). As with the Rb model, MC modeling performance is severely affected by the number of channels while the DA algorithm is independent of it and only affected by the integration time step. However, in this case MC modeling turned out to be as efficient (in some cases more efficient) than DA at the lowest dt values. This is probably due to the longer time constants of the HH model (reproducing the behavior of squid axons at 6.3ºC) compared to the Rb model (mammalian ranvier node at 37ºC). In the HH model, there are fewer transitions per time step and probably when dt<1µs there are many steps in which no transition occurs, thus leaving all the computational weight to solving the membrane current equation. However as dt increases more transitions per step begin to occur and then the computational cost is dominated by the calculation of transitions rather than by the advancing of time steps.

**Accuracy of alternative DA implementations**

Two works recently proposed DA implementations that take into account subunit coupling [33,40]. Goldwyn and colleagues [33] tested the DA approach for coupled subunit originally developed by Fox [29], and solved the square root of the stochastic diffusion matrix numerically at each time step. Besides the computational cost of this approach, it demands the matrix $D$ (eq. (14)) to be always positive semi-definite to compute real valued square roots. One simple solution for this, and the one they took, is to use the steady state approximation, replacing the values of the variables by their equilibrium values. On the other hand, Linaro *et al.* [40] deduced the covariance of the noise introduced by channel fluctuations and showed that it can be reproduced by a sum of Ornstein-Uhlenbeck processes (4 for potassium channels, 7 for sodium channels) with particular time constant and variance coefficients. This noise is then added to the sodium or potassium current, respectively, that are calculated by deterministic Hodgkin-Huxley equations. Importantly, they calculate the noise coefficients using steady-state approximation.

As shown before, the use of a steady-state approximation can result in serious deviations from the explicit MC modeling because the fluctuations become independent on the actual value of the variables at the corresponding time. Figure 10 shows that indeed this is the case, with both algorithms falling short of reproducing the behavior of Markov Chains in the voltage-clamp simulations (note the resemblance of Figure 10A with Figure 3) as well as in the firing efficiency and firing time variance curves of the Ranvier Node model (Figure 10B). We managed to implement Fox's equations without the steady-state approximation, just by extracting the absolute value of the variable vector prior to the matrix square root operation. In that case, the simulations give the same results as MC modeling and our DA implementation (not shown). Therefore, the matrix equations originally proposed by Fox and Lu are indeed a good numerical approximation to MC modeling although with a high computation cost – at least 20 times slower than our method in cases we examined.

# DISCUSSION

## Accuracy of the Diffusion Approximation

The original description of the Diffusion Approximation (DA), in its general form for a multiple (more than 2) state Markov Chain (MC), implies the calculation of the square root of a matrix [29,31]. As this is too time consuming an operation to be performed in real time, the uncoupled subunits approximation, consisting a stochastic form of the original Hodgkin and Huxley's equations, seemed to be the right choice. Very recently is was described [35] and mathematically proven [32,33,40] that when the activation subunits are considered to be coupled or 'tied' in groups (as they really are in ion channels), the resulting conductance fluctuations have statistics that cannot be adequately reproduced in a model with uncoupled subunits, either with MC modeling or a DA algorithm. Previous reports, which suggested the inadequacy of DA methods [28,35,41], failed to notice that they were using an uncoupled subunits approach for the DA and a coupled subunits approach for the MC modeling.

The simulations performed here confirm the results of Goldwyn et al. [33] in showing that the DA method was not being implemented properly for channels of more than two states. Furthermore, it is again confirmed here that the DA and MC algorithms give similar results – with two different models of neuronal excitability, and in both the coupled and uncoupled subunits approach; to our knowledge the most thorough testing that any DA algorithm has been subjected to.

## Relation to other Algorithms

Goldwyn *et al.* [33], tested the DA approach for coupled subunit originally developed by Fox [29] and showed that a properly implemented DA can approach better the results of MC modeling, in the context of the HH model. However, they computed the square root of the diffusion matrix during execution, resulting in a slow computation speed. Another recent work [40] suggested an alternative DA implementation for the HH model, that uses similar equations to the uncoupled subunits approach (Eq. (25) for $m$, $h$ and $n$) but with a noise term that is time-correlated in the way it should be when the subunits are considered to be coupled. The correlation of the noise terms requires solving 7 (Na) or 4 (K) additional differential equations, of a complexity comparable to those presented here. Importantly, both works, as well as many others, employed a steady-state approximation for the calculation of the stochastic term matrix introduced. As we showed here, this caused significant deviations in voltage clamp and the Rb model (but not in the HH model (Figure 10)). It is important to note that among the channels that work on the time scale of action potentials, the sodium channel of the Rb model has fast kinetics (resembling channels from mammalian Ranvier nodes), while the HH model possesses channels that are rather slow (giant squid axon at 6.3ºC). Most likely, this is the reason why the Rb model is more affected by the steady-state approximation than the HH model. As the time scale relevant for models based in the mammalian nervous system is precisely that of the Rb model, our conclusions about the steady-state approximation are of importance for such models.

Both previous works [33,40], as well as the original derivation by Fox [31], give specific instructions on how to construct the SDE for sodium and potassium channels, in the context of the HH model. However, generalizing these instructions to other kinetic schemes is not an easy task, since no explicit general expressions are given. In contrast, our alternative derivation gave simple and general closed form expressions for the both the diffusion matrix $D$ (eq. (10)) and its matrix square root $S$ (eq. (13)). In order to compare with previous DA formulations [31,33], we analytically found $S$ for the potassium case using Cholesky decomposition. Surprisingly, but in tune with our proposed equations, the resulting matrix was simpler (compare eq. (14) with (15)) and sparse (containing many zero elements). The exact and simple expression for $S$ (eq. (13)) allowed us to avoid the use of the inaccurate steady state approximation and to improve simulation speed considerably. Specifically, instead of the $O(M^3)$ computational complexity of the numerical matrix square root implementation (as done in [33]) our method has a complexity between $O(M)$ and $O(M^2)$, depending on the number of kinetic transitions (see eq. (13)). Numerically testing this, we observed our method run at least 20 times faster, depending on the software environment employed. Also our method only doubles the time required to solve the deterministic equations, that ignore the stochastic terms (not shown). Moreover, the equations that govern the dynamics of stochastic ion channels in our approach can be simply written as separate equations instead of matrix operations (e.g. eq. (16) for potassium and Supplemental Text T1 for sodium). This facilitates their implementation in non matrix-oriented computation software such as Neuron, and may also simplify future analytical analysis of the behavior of the stochastic neuron.

We note a connection between the DA approach and another stochastic simulation method - the "binomial population" approach [46,47,48]. This approach employs eq. (11) directly, where each $\Delta_{ij}$ is distributed binomially. So essentially, the only additional approximation we made was that $\Delta_{ij}$ was a Gaussian RV. This can greatly reduce simulation speed since the generation of binomial RVs is much less efficient than Gaussian RVs, especially for large $N$ [49]. As noted, our simulations (as well as Goldwyn's [33] and Linaro's [40]) indicate that this approximation is very good, as long as $N$ is not too small. However, if $N$ is small enough, so that the discrete nature of ion channel conductance becomes significant, then this approximation might break down. In that case, one can speculate that it might be more accurate to approximate $\Delta_{ij}$ as a Poisson RV with parameter $np = X_j A_{ij} dt$ (by the law of rare events). Note that also in this approximation it is possible to pair opposing transition pairs $W_{ij} = \Delta_{ij} - \Delta_{ji}$ (as in eq. (12) ) and generate $W_{ij}$ according to a Skellam distribution (the distribution of the difference between two Poisson RV). However, we have not investigated here whether or not the Poisson\Skellam approximations may actually improve the speed of binomial population algorithm or have any advantage over other methods (such as MC).

Finally, we note that a similar approach to ours was previously introduced in the field of chemical physics. As in our case, this equation, named "the *chemical Langevin equation*" [50,51] sums the stochastic terms along transitions and not along states (compare our eq. (13) with eq. (23) in [50]).

The main difference between that approach and ours is that we sum together the noise contributions from both directions of each transition pair (done in the conversion from eq. (11) to eq. (12)). This approximately halves the computation time, when the generation of pseudo-RVs is the main computational bottleneck.

**Numerical efficiency – DA versus MC**

Following the practical approach of this work, we numerically evaluated the computational cost of both implementations (MC and DA). In almost all cases, our DA approach significantly outperforms the MC approach. In the Rb simulations (Figure 6C&D) the DA approach for coupled subunits is at least an order of magnitude faster than MC for all values of $N$ and $dt$ tested. In the HH simulations (Figure 9D&E) this remains true, except when low values of $dt$ or $N$ are used. Again we note that the results for Rb model are more significant to the mammalian nervous system, due to the similar kinetic timescales. Another issue to consider when comparing Figures 6 and 9 is that the Rb simulations presented here were performed in the Scilab numerical computation package while the HH simulations were implemented in NEURON. The latter will be always faster because it runs as compiled code; also variations in how each software implements numerical calculations at the processor level may cause further differences.

In all cases, however, the speed of simulations performed with the DA algorithm was only affected by the size of the integration time step and completely independent of the number of channels to be simulated, because the number of channels is only a parameter in the equations. On the contrary, MC modeling was heavily affected by the number of channels and less affected by the integration time step. In this case a greater number of channels imply more transitions per time step, and for each transition two new calculations have to be made, each requiring a new random number.

Thus, there will be situations where MC modeling may be numerically more efficient than DA. With a small number of channels there will be fewer transitions per time step and thus a MC simulation may run faster than a DA algorithm. This difference will be enhanced if the channels have slow kinetics, because this will reduce the probability of transitions. Also, if a small integration time step is required the DA algorithm can be as slow as MC modeling. In both these cases, it might be better to combine the MC and DA methods: use MC for channel with slow kinetics, while handle the faster channels using the DA approach. The waterline between "slow" and "fast" timescales here would be the time step duration. Also, note that in the simulations presented here, random numbers were generated in simulation time. Further speed-up of the DA algorithm can be achieved by the use of a pre-generated random number list.

**Conclusions**

This paper further confirms that the use of the Diffusion Approximation (DA), without any additional approximations, produce results that are indistinguishable from those of Markov Chain modeling (MC). Most importantly, we present the DA in a very simple, general and computationally efficient form, which will allow its easy implementation for any given kinetic scheme of a channel. We show that in the most common situations, the DA method proposed here has a numerical stability comparable to that of MC modeling (even with a simple Euler-Maruyama integration scheme), while being much faster. The fast simulation speed achieved makes conceivable its use in dynamic clamp experiments.

# METHODS

## Models

To test the accuracy and efficiency of DA relative to MC modeling, both in their independent subunits and coupled subunits approaches, two models were employed in which different measures of simulation accuracy were calculated.

*Mammalian Ranvier node – Rb model*

The mammalian Ranvier node model [43] was the model employed previously to compare the performance of DA versus MC modeling [28,41]. This model consists only of a voltage-dependent sodium channel and a voltage-independent leak current. The membrane current equation is

$$C_m \frac{dV(t)}{dt} + \frac{V(t)}{R} + g_{Na}(t)\left(V(t) - E_{Na}\right) = I_{app}(t) \tag{18}$$

with parameters $C_m$=18.9 pF; $R$=7.372 MΩ; $\overline{g_{Na}}$ =6.808 µS; $E_{Na}$=144 mV. The voltage is shifted so that the leak reversal potential is 0. The $\alpha$ and $\beta$ transition rates are given by the following voltage dependent functions:

$$\begin{aligned}
\alpha_m(V) &= \frac{1.872(V - 25.41)}{1 - \exp\left((25.41 - V)/6.06\right)} \\
\beta_m(V) &= \frac{3.973(21 - V)}{1 - \exp\left((v - 21)/9.41\right)} \\
\alpha_h(V) &= -\frac{0.549(27.74 + V)}{1 - \exp\left((V + 27.74)/9.06\right)} \\
\beta_h(V) &= \frac{22.57}{1 + \exp\left((56 - V)/12.5\right)}
\end{aligned} \tag{19}$$

Simulations of 1 ms were run in which a 100 µs current pulse was given at the beginning (Figure 1). The pulse amplitude varied between 5 and 6.5 pA. 1,000 simulations were run and the following parameters were calculated: *Firing efficiency*, the fraction of simulations in which an action potential was evoked; and the mean and the variance of *Firing time,* time at which the voltage reached or surpassed 80 mV. Firing efficiency versus pulse amplitude curve was fit to the cumulative Gaussian distribution

$$FiringEfficiency\left(I_{app}\right) = \Phi\left(\frac{I_{app} - Th}{\sigma}\right)$$

$$\Phi(x) = \frac{1}{\sqrt{2\pi}} \int_{-\infty}^{x} e^{\frac{-t^2}{2}} dt = \frac{1}{2}\left[1 + erf\left(\frac{x}{\sqrt{2}}\right)\right]$$

*erf(x)* represents the error function. *Th (threshold)* gives the amplitude for a probability of firing of 0.5, while $\sigma$ quantifies the spread of the input/output relationship.

*Hodgkin and Huxley model of squid giant axon – HH model*

The original Hodgkin and Huxley [21] model was simulated with the equation

$$C_m \frac{dV(t)}{dt} = -g_{Na}(t)\left(V(t) - E_{Na}\right) - g_K(t)\left(V(t) - E_K\right) - g_l\left(V(t) - E_l\right) \quad (20)$$

and parameters $C_m=1$, $E_{Na}=50$ mV, $E_K=-77$ mV, $E_l=-54.4$ mV, $\overline{g_{Na}}=120$, $\overline{g_K}=36$, $g_l=0.3$ (Voltages are shifted with respect to the original model to make the resting potential equal to -65 mV). The α and β functions employed are

$$\alpha_m(V) = \frac{0.1(V+40)}{1-\exp\left(-\frac{V+40}{10}\right)} \quad \beta_m(V) = 4\exp\left(-\frac{V+65}{18}\right)$$

$$\alpha_h(V) = 0.07\exp\left(-\frac{V+65}{20}\right) \quad \beta_h(V) = \frac{1}{1+\exp\left(-\frac{V+35}{10}\right)} \quad (21)$$

$$\alpha_n(V) = \frac{0.01(V+55)}{1-\exp\left(-\frac{V+55}{10}\right)} \quad \beta_n(V) = 0.125\exp\left(-\frac{V+65}{80}\right)$$

Simulations of 500 seconds were performed, and action potentials were recorded as the time at which the voltage reached or surpassed 0 mV. The time of action potentials during the simulation were stored, and the Inter-Spike Intervals (ISIs) were calculated. The normalized ISI distribution was fitted to an exponential decay function with a refractory period [16]:

$$P(\tau) = r\exp\left(-r(\tau - t_{ref})\right) \quad (22)$$

The first two values of the ISI distribution histogram were not included in the fitting procedure.

**Uncoupled independent subunits**

$N$ channels are simulated as $4N$ independent, 2-state subunits:

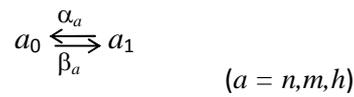

$$(a = n, m, h)$$

where $\alpha_a$ is the transition probability from the 0 to the 1 state, and $\beta_a$ the transition probability from the 1 to the 0 state. $N_{Na}$ Sodium channels are simulated as $3N_{Na}$ $m$ subunits and $N_{Na}$ $h$ subunits, and at each time step the sodium conductance is calculated as

$$g_{Na} = \overline{g_{Na}}\left(\frac{Nm_1}{3N_{Na}}\right)^3\left(\frac{Nh_1}{N_{Na}}\right). \quad (23)$$

$N_K$ Potassium channels are simulated as $4N_K$ $n$ subunits and the potassium conductance is calculated as

$$g_K = \overline{g_K}\left(\frac{Nn_1}{4N_K}\right)^4. \quad (24)$$

$Nm_1$, $Nh_1$ and $Nn_1$ are the number of $m$, $h$, and $n$ subunits, respectively, that are in the '1' state.

*Diffusion Approximation*

The DA in the case of independent subunits uses the variables $m$, $h$, $n \in [0,1]$ to keep track of the fraction of $m$, $h$, and $n$ subunits, respectively, that are in the '1' state. It follows immediately that the fraction of subunits in the '0' state will be 1-$m$, 1-$h$, and 1-$n$, respectively. Fox and Lu [29] showed that the time evolution of the variables is given by the SDE

$$\frac{da}{dt} = \alpha_a(1-a) - \beta_a a + \sigma_a(t)\xi(t) \tag{25}$$

where $a$ represents either $m$, $h$ or $n$. The stochastic term $\xi(t)$ is a Gaussian white noise with zero mean and unit variance that is scaled by $\sigma_a(t)$, being

$$\sigma_a(t) = \sqrt{\frac{\alpha_a(1-a) + \beta_a a}{N_a}} \tag{26}$$

where $N_a$ is the number of $a$ subunits ($N_m = 3N_{Na}$, $N_h = N_{Na}$ and $N_a = 4N_K$). When the steady state approximation was used, the noise scaling factor was calculated as

$$\sigma_a(t) = \sqrt{\frac{2\alpha_a \beta_a}{N_a(\alpha_a + \beta_a)}} \tag{27}$$

The conductance of sodium and potassium are calculated using the classical Hodgkin & Huxley expressions

$$g_{Na} = \overline{g_{Na}} m^3 h \quad \text{and} \quad g_K = \overline{g_K} n^4.$$

In the voltage clamp simulations, the number of open potassium channels was calculated as

$$N_O = N_K n^4.$$

## Coupled independent subunits (independent channels)

There are two possible ways of implementing a coupled subunits approach. The first consist of simulating 4 independent 2-state subunits per channel, as in the previous approach. However, each subunit is assumed to belong to a specific channel. A channel is considered open if and only if its four subunits are in the open state. Therefore, the state of each subunit (hence of each channel) must be tracked individually during the simulation [43].

In this paper a second approach is employed, that consists in building a multi-state MC per channel considering the possible combinations of active subunits. This allows for the faster number-tracking algorithm employed for simulations [16,26,27]. Given that subunits of a given kind are identical and independent, a Sodium channel has 8 possible states while a Potassium channel has 5 states:

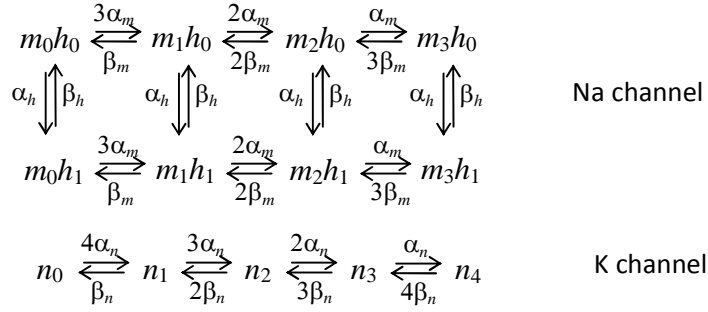

Na channel

K channel

In this approach, only one state of each MC represents the conducting or open channel, which is the state with all subunits active ($m_3h_1$ or $n_4$). Then the conductance is calculated with the fraction of channels or MCs that are in the open state:

$$g_{Na} = \overline{g_{Na}}\left(\frac{Nm_3h_1}{N_{Na}}\right); g_K = \overline{g_K}\left(\frac{Nn_4}{N_K}\right)$$

where $Nm_3h_1$ and $Nn_4$ are the number of channels in the state $m_3h_1$ and $n_4$, respectively.

*Diffusion Approximation*

The DA for channels with coupled activation subunits is detailed in the *Results* section.

## Numerical implementations

*Software implementation*

All models and algorithms were implemented in Scilab, a matrix-oriented numerical software (www.scilab.org), and NEURON, a simulation environment oriented to the modeling neurons and neural networks (www.neuron.yale.edu). Source files and scripts are available in ModelDB (http://senselab.med.yale.edu/ModelDB/). Both environments produced identical results but simulations in NEURON run faster because it runs in compiled mode. Results presented here (most importantly, processing time data) correspond to simulations in Scilab for the mammalian Ranvier node (Rb) model and simulations in NEURON for the squid giant axon (HH) model.

*Markov Chain modeling*

Independent MCs were modeled using a number-tracking algorithm [16,26,27,28]. Thoroughly described in [28], briefly this algorithm consist in keeping track of the number of MCs in each state, rather than keeping track of each MC individually. At any time $t$, the probability density function of the lifetime before the next transition (any transition) is

$$P_t(\tau) = \lambda(t)\exp(-\lambda(t)\tau)$$

where $\lambda(t)$ is the effective transition rate given by

$$\lambda(t) = \sum_i^S N_i(t)\zeta_i(t)$$

where $S$ is the total number of states in the MC, $N_i$ is the number of MCs in state $i$, and $\zeta(t)$ is the sum of transition rates escaping from state $i$. If there is more than one type of MC, they are all summed into $\lambda$. The time of the next transition $t_n$ is calculated by drawing a random number uniformly distributed within [0,1] and taking the inverse of the c.d.f. of the lifetime. If $t_n \leq t$, a transition has to be calculated before updating the current equation. Among all possible transitions, the probability of transition *j* to occur is

$$P_j(t) = N_i(t)\alpha_j(t)$$

where *i* is the state originating the transition *j* and $\alpha_j$ its rate. A cumulative probability for all transitions is calculated and a transition is chosen by drawing a random number uniformly distributed within [0,1]. The number of MCs at each state is updated, and a new time for the next transition is calculated. When no more transitions are to occur in the current time step, the current equation is advanced one time step using an Euler integration scheme.

*Diffusion approximation*

Stochastic differential equations for DA were solved by an Euler-Maruyama integration method. For the coupled subunits approach, a better numerical stability is obtained if the fact that the sum of state variables for a given channel is 1 is taken into account, also reducing the number of SDEs to be solved. Thus, for potassium channels the equations used for advancing one time step are

$$n_{1,t+dt} = n_1 + dt\left(4\alpha_n n_0 - \beta_n n_1 - 3\alpha_n n_1 + 2\beta_n n_2\right) - \eta_1\sqrt{4\alpha_n n_0 + \beta_n n_1} + \eta_2\sqrt{3\alpha_n n_1 + 2\beta_n n_2}$$

$$n_{2,t+dt} = n_2 + dt\left(3\alpha_n n_1 - 2\beta_n n_2 - 2\alpha_n n_2 + 3\beta_n n_3\right) - \eta_2\sqrt{3\alpha_n n_1 + 2\beta_n n_2} + \eta_3\sqrt{2\alpha_n n_2 + 3\beta_n n_3}$$

$$n_{3,t+dt} = n_3 + dt\left(2\alpha_n n_2 - 3\beta_n n_3 - \alpha_n n_3 + 4\beta_n n_4\right) - \eta_3\sqrt{2\alpha_n n_2 + 3\beta_n n_3} + \eta_4\sqrt{\alpha_n n_3 + 4\beta_n n_4}$$

$$n_{4,t+dt} = n_4 + dt\left(\alpha_n n_3 - 4\beta_n n_4\right) - \eta_4\sqrt{\alpha_n n_3 + 4\beta_n n_4}$$

$$n_{0,t+dt} = 1 - n_{1,t+dt} - n_{2,t+dt} - n_{3,t+dt} - n_{4,t+dt}$$

being $\eta_1$, $\eta_2$, $\eta_3$, and $\eta_4$ independent Gaussian RVs with zero mean and standard deviation $\sqrt{dt/N_K}$. $n_0 - n_4$ stand for $n_{0,t} - n_{4,t}$, i.e. the value of the variables at time *t*. A similar set of equations was used for sodium channels.

No rounding was performed on the variables, nor were they bound to lie between 0 and 1 (see discussion). To ensure real valued random terms, the square roots were applied to the absolute value of the operand.

For the steady state approximation, the variables $n_i$ and $m_i h_j$ were replaced by their steady state values in all the noise terms:

$$n_i^\infty = \binom{4}{i}\frac{\alpha_n^i \beta_n^{4-i}}{(\alpha_n + \beta_n)^4}, \; m_i h_j^\infty = \binom{3}{i}\frac{\alpha_m^i \beta_m^{3-i}\alpha_h^j \beta_h^{1-j}}{(\alpha_m + \beta_m)^3(\alpha_h + \beta_h)}. \quad (28)$$

## ACKNOWLEDGMENTS

We wish to thank Rolando Rebolledo, Mauricio Tejo and Rolando Biscay for helpful discussions and Joshua Goldwyn, Yuval Elhanati, Rolando Rebolledo and Ron Meir for critical reading of the manuscript.
27

# REFERENCES

1. Faisal AA, Selen LP, Wolpert DM (2008) Noise in the nervous system. Nat Rev Neurosci 9: 292-303.
2. Stein RB, Gossen ER, Jones KE (2005) Neuronal variability: noise or part of the signal? Nat Rev Neurosci 6: 389-397.
3. Longtin A, Bulsara A, Moss F (1991) Time-interval sequences in bistable systems and the noise-induced transmission of information by sensory neurons. Phys Rev Lett 67: 656-659.
4. Gluckman BJ, Netoff TI, Neel EJ, Ditto WL, Spano ML, et al. (1996) Stochastic Resonance in a Neuronal Network from Mammalian Brain. Phys Rev Lett 77: 4098-4101.
5. McDonnell MD, Abbott D, Friston KJ (2009) What Is Stochastic Resonance? Definitions, Misconceptions, Debates, and Its Relevance to Biology. PLoS Comput Biol 5: e1000348.
6. McDonnell MD, Ward LM (2011) The benefits of noise in neural systems: bridging theory and experiment. Nature reviews Neuroscience 12: 415-426.
7. Douglass JK, Wilkens L, Pantazelou E, Moss F (1993) Noise enhancement of information transfer in crayfish mechanoreceptors by stochastic resonance. Nature 365: 337-340.
8. Levin JE, Miller JP (1996) Broadband neural encoding in the cricket cercal sensory system enhanced by stochastic resonance. Nature 380: 165-168.
9. Molgedey L, Schuchhardt J, Schuster HG (1992) Suppressing chaos in neural networks by noise. Physical review letters 69: 3717-3719.
10. Fiete IR, Seung HS (2006) Gradient learning in spiking neural networks by dynamic perturbation of conductances. Physical review letters 97: 048104.
11. Kirkpatrick S, Gelatt CD, Jr., Vecchi MP (1983) Optimization by simulated annealing. Science 220: 671-680.
12. Motwani R, Raghavan P (1996) Randomized algorithms. ACM Comput Surv 28: 33-37.
13. Lecar H, Nossal R (1971) Theory of threshold fluctuations in nerves. II. Analysis of various sources of membrane noise. Biophys J 11: 1068-1084.
14. White JA, Rubinstein JT, Kay AR (2000) Channel noise in neurons. Trends Neurosci 23: 131-137.
15. White JA, Klink R, Alonso A, Kay AR (1998) Noise from voltage-gated ion channels may influence neuronal dynamics in the entorhinal cortex. J Neurophysiol 80: 262-269.
16. Chow CC, White JA (1996) Spontaneous action potentials due to channel fluctuations. Biophys J 71: 3013-3021.
17. Faisal AA, Laughlin SB (2007) Stochastic Simulations on the Reliability of Action Potential Propagation in Thin Axons. PLoS Comput Biol 3: e79.
18. Rowat PF, Elson RC (2004) State-dependent effects of Na channel noise on neuronal burst generation. J Comput Neurosci 16: 87-112.
19. Dorval AD, Jr., White JA (2005) Channel noise is essential for perithreshold oscillations in entorhinal stellate neurons. J Neurosci 25: 10025-10028.
20. Fernandez FR, White JA (2008) Artificial Synaptic Conductances Reduce Subthreshold Oscillations and Periodic Firing in Stellate Cells of the Entorhinal Cortex. J Neurosci 28: 3790-3803.
21. Hodgkin AL, Huxley AF (1952) A quantitative description of membrane current and its application to conduction and excitation in nerve. J Physiol 117: 500-544.
22. Hille B (2001) Ion Channels of Excitable Membranes. Sunderland, MA, USA: Sinauer Associates Inc.
23. Bezanilla F (2000) The voltage sensor in voltage-dependent ion channels. Physiol Rev 80: 555-592.

24. Colquhoun D, Hawkes AG (1981) On the stochastic properties of single ion channels. Proc R Soc Lond B Biol Sci 211: 205-235.
25. Neher E, Stevens CF (1977) Conductance fluctuations and ionic pores in membranes. Annu Rev Biophys Bioeng 6: 345-381.
26. Gillespie DT (1977) Exact stochastic simulation of coupled chemical reactions. J Phys Chem 81: 2340-2361.
27. Skaugen E, Walloe L (1979) Firing behaviour in a stochastic nerve membrane model based upon the Hodgkin-Huxley equations. Acta Physiol Scand 107: 343-363.
28. Mino H, Rubinstein JT, White JA (2002) Comparison of algorithms for the simulation of action potentials with stochastic sodium channels. Ann Biomed Eng 30: 578-587.
29. Fox RF, Lu Y (1994) Emergent collective behavior in large numbers of globally coupled independently stochastic ion channels. Phys Rev E Stat Phys Plasmas Fluids Relat Interdiscip Topics 49: 3421-3431.
30. Goldwyn JH, Shea-Brown E (2011) The what and where of adding channel noise to the Hodgkin-Huxley equations. PLoS Comput Biol: (in press).
31. Fox RF (1997) Stochastic versions of the Hodgkin-Huxley equations. Biophys J 72: 2068-2074.
32. Pakdaman K, Thieullen M, Wainrib G (2010) Fluid Limit Theorems for Stochastic Hybrid Systems with Application to Neuron Models. Adv in Appl Probab 42: 761-794.
33. Goldwyn JH, Imennov NS, Famulare M, Shea-Brown E (2011) Stochastic differential equation models for ion channel noise in Hodgkin-Huxley neurons. Phys Rev E Stat Nonlin Soft Matter Phys 83: 041908.
34. Golub G, van Loan C (1996) Matrix Computations (Johns Hopkins Studies in Mathematical Sciences)(3rd Edition): The Johns Hopkins University Press.
35. Bruce IC (2009) Evaluation of stochastic differential equation approximation of ion channel gating models. Ann Biomed Eng 37: 824-838.
36. Schoppa NE, Sigworth FJ (1998) Activation of Shaker potassium channels. III. An activation gating model for wild-type and V2 mutant channels. J Gen Physiol 111: 313-342.
37. Bezanilla F, Perozo E, Stefani E (1994) Gating of Shaker K+ channels: II. The components of gating currents and a model of channel activation. Biophys J 66: 1011-1021.
38. Horn R, Ding S, Gruber HJ (2000) Immobilizing the moving parts of voltage-gated ion channels. J Gen Physiol 116: 461-476.
39. Vandenberg CA, Bezanilla F (1991) A sodium channel gating model based on single channel, macroscopic ionic, and gating currents in the squid giant axon. Biophys J 60: 1511-1533.
40. Linaro D, Storace M, Giugliano M (2011) Accurate and fast simulation of channel noise in conductance-based model neurons by diffusion approximation. PLoS Comput Biol 7: e1001102.
41. Bruce IC (2007) Implementation issues in approximate methods for stochastic Hodgkin-Huxley models. Ann Biomed Eng 35: 315-318; author reply 319.
42. Alvarez O, Gonzalez C, Latorre R (2002) Counting channels: a tutorial guide on ion channel fluctuation analysis. Adv Physiol Educ 26: 327-341.
43. Rubinstein JT (1995) Threshold fluctuations in an N sodium channel model of the node of Ranvier. Biophys J 68: 779-785.
44. Mauro A, Conti F, Dodge F, Schor R (1970) Subthreshold behavior and phenomenological impedance of the squid giant axon. J Gen Physiol 55: 497-523.
45. Koch C (1999) Biophysics of computation : information processing in single neurons. Oxford: Oxford University Press. xxiii, 562 p. p.
46. Schneidman E, Freedman B, Segev I (1998) Ion channel stochasticity may be critical in determining the reliability and precision of spike timing. Neural Comput 10: 1679-1703.

47. Steinmetz PN, Manwani A, Koch C, London M, Segev I (2000) Subthreshold voltage noise due to channel fluctuations in active neuronal membranes. J Comput Neurosci 9: 133-148.
48. Faisal A (2010) Stochastic Simulation of Neurons, Axons, and Action Potentials. In: Laing C, Lord GJ, editors. Stochastic Methods in Neuroscience. New York: Oxford University Press.
49. Press WH (2007) Numerical recipes : the art of scientific computing. Cambridge, UK ; New York: Cambridge University Press. xxi, 1235 p. p.
50. Gillespie DT (2000) The chemical Langevin equation. J Chem Phys 113: 297.
51. Bhalla US, Wils S (2010) Reaction-Diffusion Modeling. In: De Schutter E, editor. Computational Modeling Methods for Neuroscientists. Cambridge, Massachusetts, Londond, England: The MIT Press.

# FIGURE LEGENDS

**Figure 1. Comparison between models with uncoupled activation subunits and coupled activation subunits.**

**Figure 2. Voltage clamp simulation and non-stationary noise analysis.**
300 potassium channels from the HH model were simulated at a constant voltage of 70 mV. At t=0, they were in a steady state condition calculated at -90 mV. 200 independent simulations were performed with each simulation algorithm (indicated above each panel) and a non-stationary noise analysis was performed. Top row: 8 sample traces of the number of open channels against time. Middle: Mean (black) and variance (grey) of number of open channels as a function of time. Bottom: The variance of the number of open channels is plotted against the mean. The continuous line represents the best fit to equation (17), and the best fit parameters are indicated below.

**Figure 3. Effect of the steady-state approximation on the voltage clamp simulations.**
The same simulations as in Figure 1 were performed with the DA method, however the value of the variables in the random terms were replaced by their steady-state values (eq. (28)). The top, middle and bottom panels are as in Figure 1.

**Figure 4. Rb model simulations.**
**A.** 15 voltage traces (bottom) resulting from independent simulations with the Rb model, in which a 5.8 pA pulse of 100 µs duration (top) was applied. The simulations presented correspond to the Rb8 model (independent channels approach) using MC modeling, with 1000 Na channels and dt=1 µs. **B-D.** Firing efficiency (fraction of action potentials evoked in 1000 simulations), mean firing time, and variance of firing time as a function of stimulus amplitude for the different simulation methods. Rb2: independent subunits, Rb8: independent channels (tied subunits), MC: Markov chain modeling, DA: Diffusion approximation algorithm. N=1000, dt=0.1 µs

**Figure 5. Quantification of variability in the Rb model and its dependence on the number of channels simulated.**
**A.** Fitting of a firing efficiency curve to a sigmoid function (see Methods) that is characterized by a threshold (the stimulus amplitude that produces a firing efficiency of 0.5) and σ (the standard deviation of the threshold fluctuations). **B.** Dependence of the threshold and slope values on the number of channels simulated, for each of the simulation methods. dt=1 µs. **C.** Effect of the steady-state approximation (DA-ss) on the behavior of the model. Comparison of a firing efficiency curve for 1000 channels (left), and the fitting parameters of the firing efficiency (middle and right) for simulations performed with DA algorithm with and without steady state approximation.

**Figure 6. Numerical stability and computational cost of the simulation algorithms with the Rb model.**
**A-B.** Dependence of Rb model variability on the integration time step used in the simulation. Threshold (**A**) and σ (**B**) values calculated as in figure 4A, obtained at different values of integration time step (dt). N=1000. **C.** Dependence of computation time on integration time step (dt) with N=1000 channels. **D.** Dependence of computation time on number of channels (N) with dt=0.5 µs. Computation time is the time, in seconds, needed to perform the 16000 simulations necessary for a single firing efficiency curve (1000 pulses at 16 current levels). This figure corresponds to simulations performed in the Scilab numerical computation software. 1000 simulations were performed as 10 batches of 100 simultaneous and independent simulations, in a Core i7 machine.

**Figure 7. Spontaneous firing in the Hodgkin and Huxley squid axon model.**
Sample voltage traces of 2 seconds of simulation of the stochastic HH model for all models and simulation algorithms tested.

**Figure 8. Firing rate and ISI distributions for the stochastic HH models.**
**A.** Mean firing rate of the stochastic HH models in a 500 seconds simulation with different number of channels. **B.** An inter-spike interval (ISI) was built for each simulation and the data was fitted to an exponential decay function with a refractory period (see Methods and ref. [16]). The histograms for only two simulations are shown here for illustration purposes. The first two points (marked with asterisks) were omitted in the fitting procedure (see text). The fit lines for the two histograms showed here overlap almost perfectly. **C-D.** Fit parameters of the ISI distributions at different number of channels. In all the simulations, $N_K=0.3*N_{Na}$. Data in this figure corresponds to dt=0.1 µs.

**Figure 9. Numerical stability and computational cost of the simulation algorithms with the HH model.**
**A-C.** Firing parameters of the stochastic HH models at different integration time steps. Mean firing rate (**A**) and fitting parameters of the ISI distributions (**B-C**) for the stochastic HH models tested as a function of the integration time step (dt). For the HH58 models, $N_{Na}=3000$ and $N_K=900$; for the HH2 models, $N_{Na}=300$ and $N_K=90$. **D.** Time to perform a 500 seconds simulation with $N_{Na}=6000$ and $N_K=1800$ as a function of dt. **E.** Time to perform a 500 seconds simulation with dt=5 µs as a function of $N_{Na}$, the number of Na channels. $N_K=0.3*N_{Na}$. The segmented line indicates the 500 seconds limit; any simulation below this line run faster than real time.

**Figure 10. Inaccuracies introduced by previous DA algorithms.**
**A.** Performance of the of the Fox [29] algorithm for coupled subunits employed by Goldwyn *et al.* [33] and the Linaro *et al.* algorithm [40] in the voltage clamp simulation and non-stationary noise analysis. See legend of Figure 2 for further details. **B.** Performance of the algorithms in the Ranvier Node model simulations. Firing Efficiency, Firing Time Variance and Mean Firing Time versus Stimulus Amplitude are presented for simulations with $N_{Na}=1000$. Standard Deviation for Threshold (σ) is plotted against number of channels (see Figure 5). dt=0.5 µs.

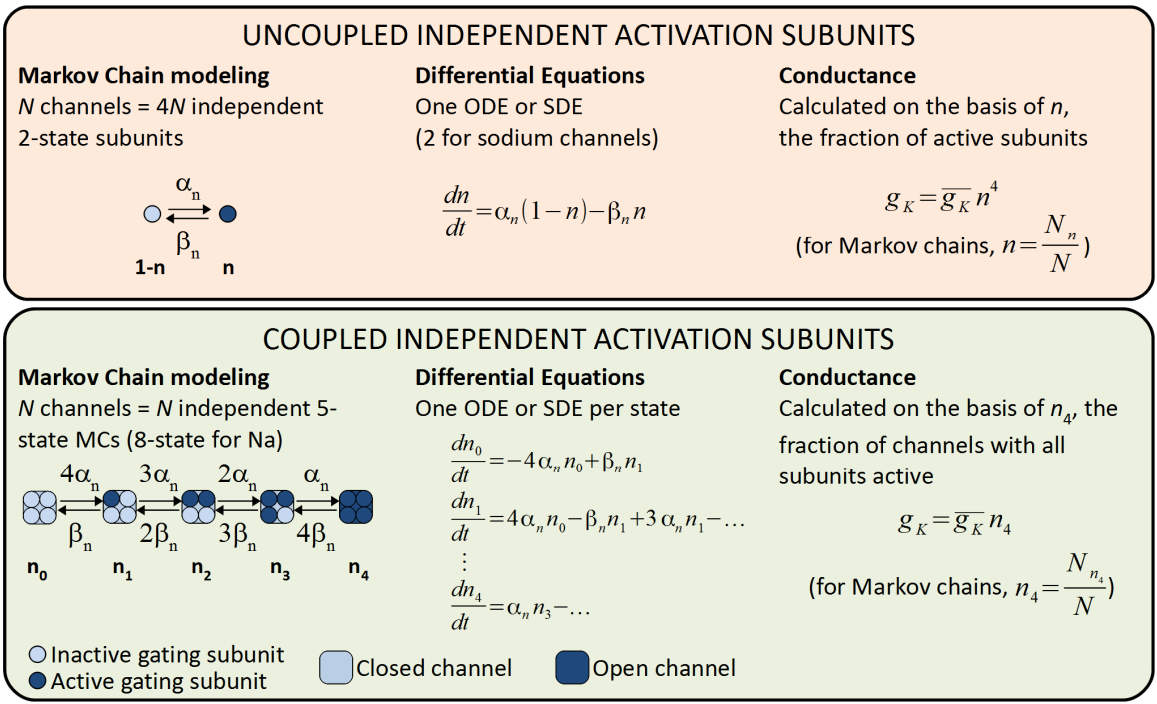

Figure 1

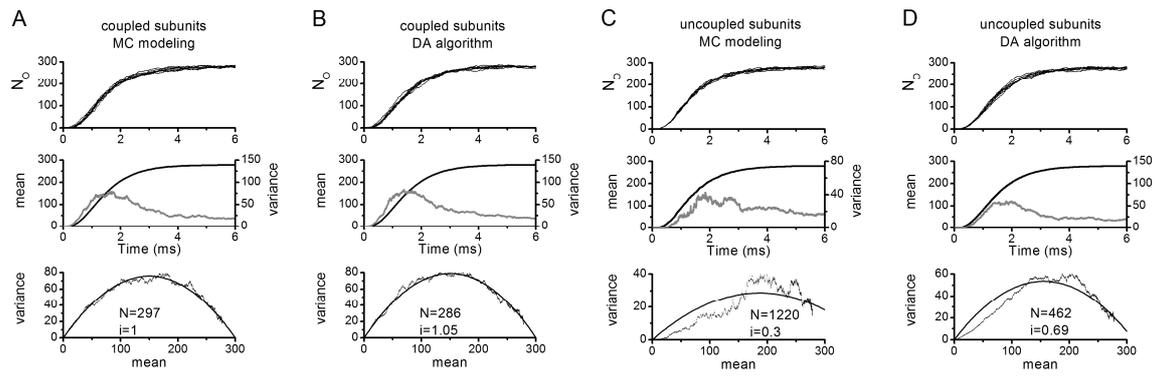

Figure 2

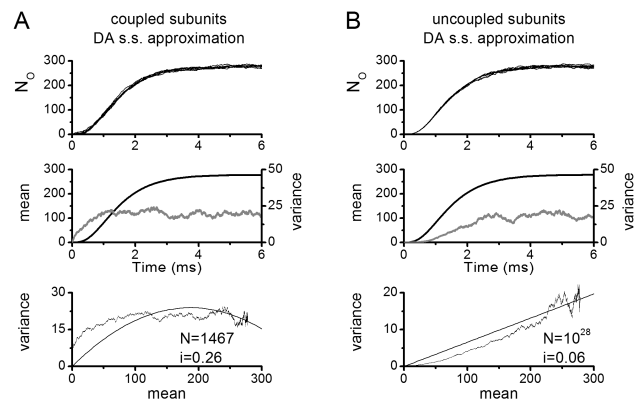

Figure 3

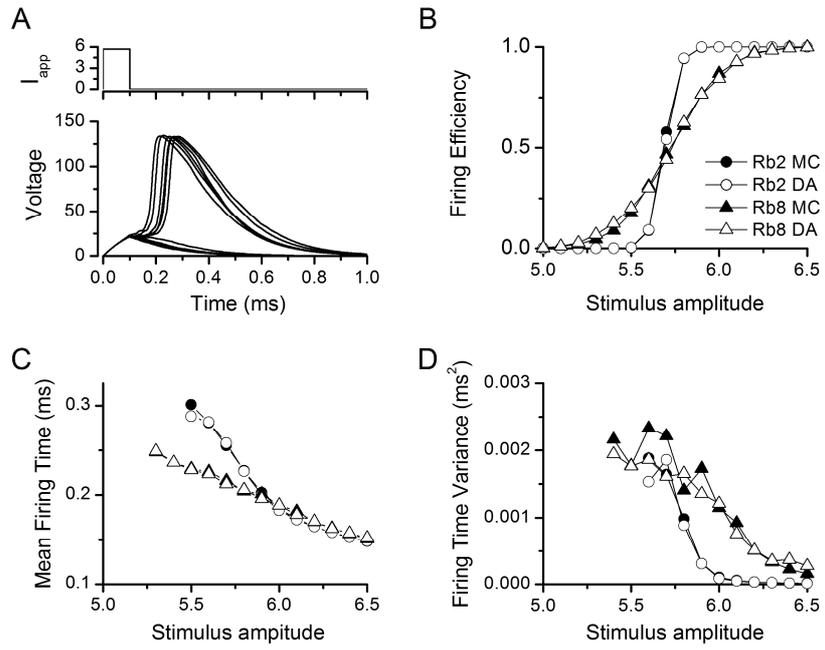

Figure 4

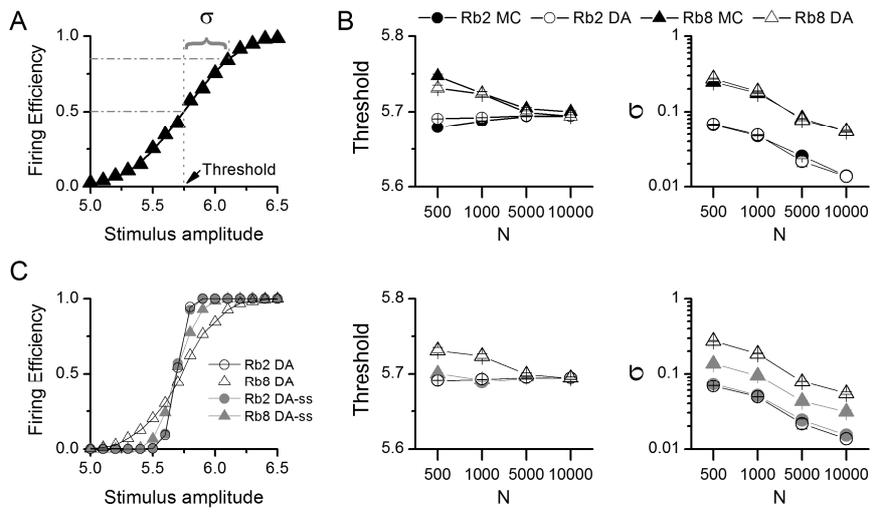

Figure 5

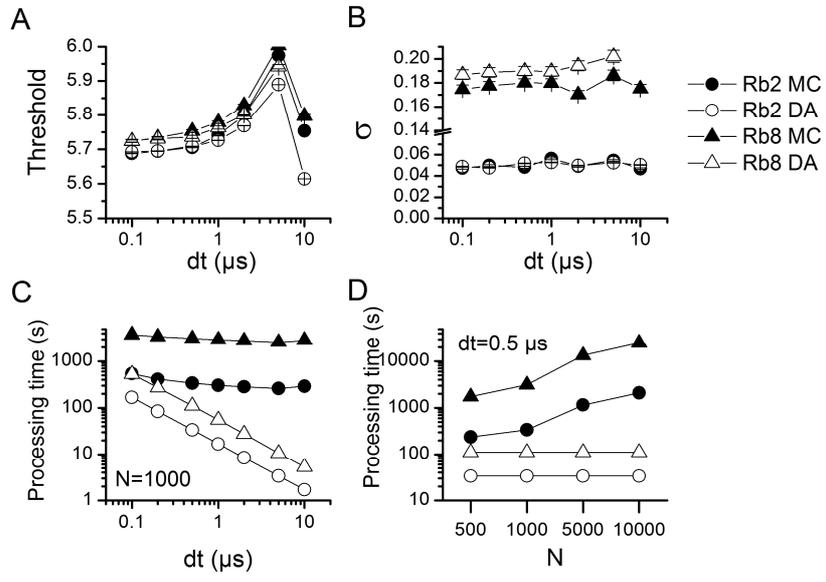

Figure 6

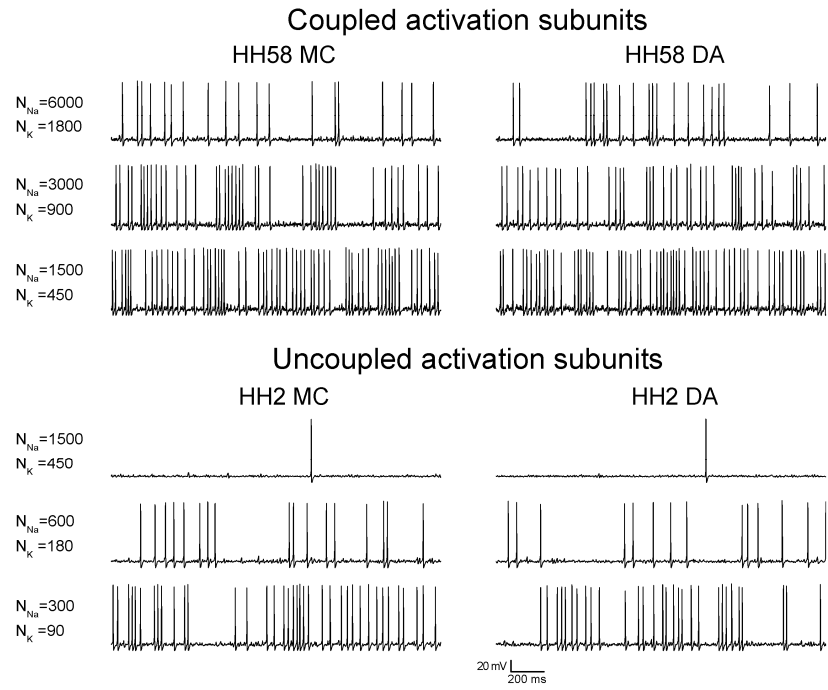

Figure 7

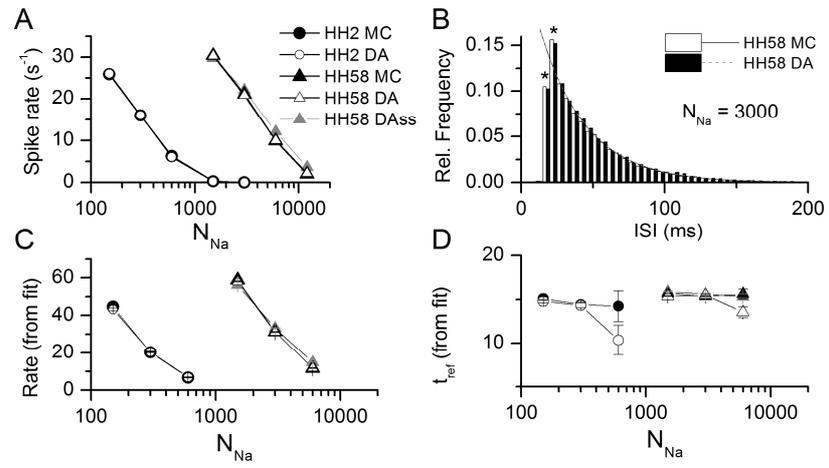

Figure 8

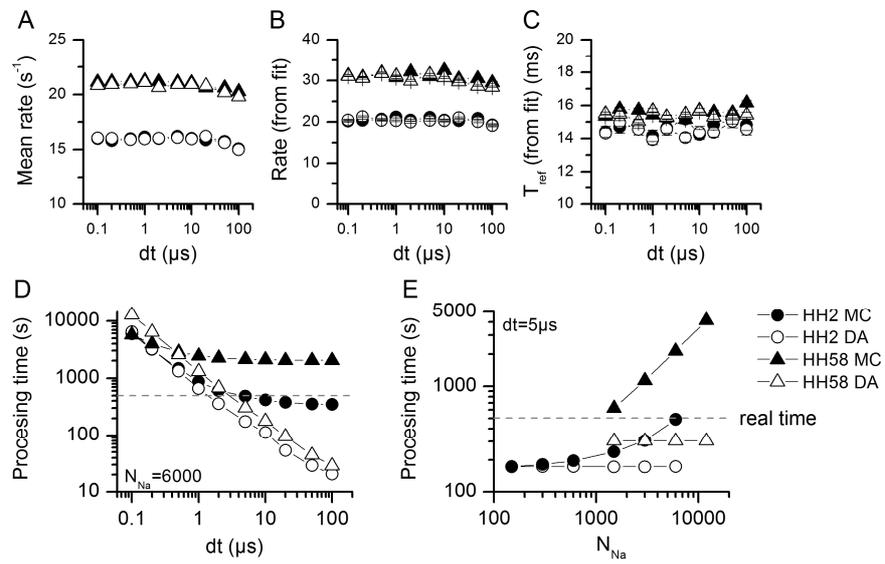

Figure 9

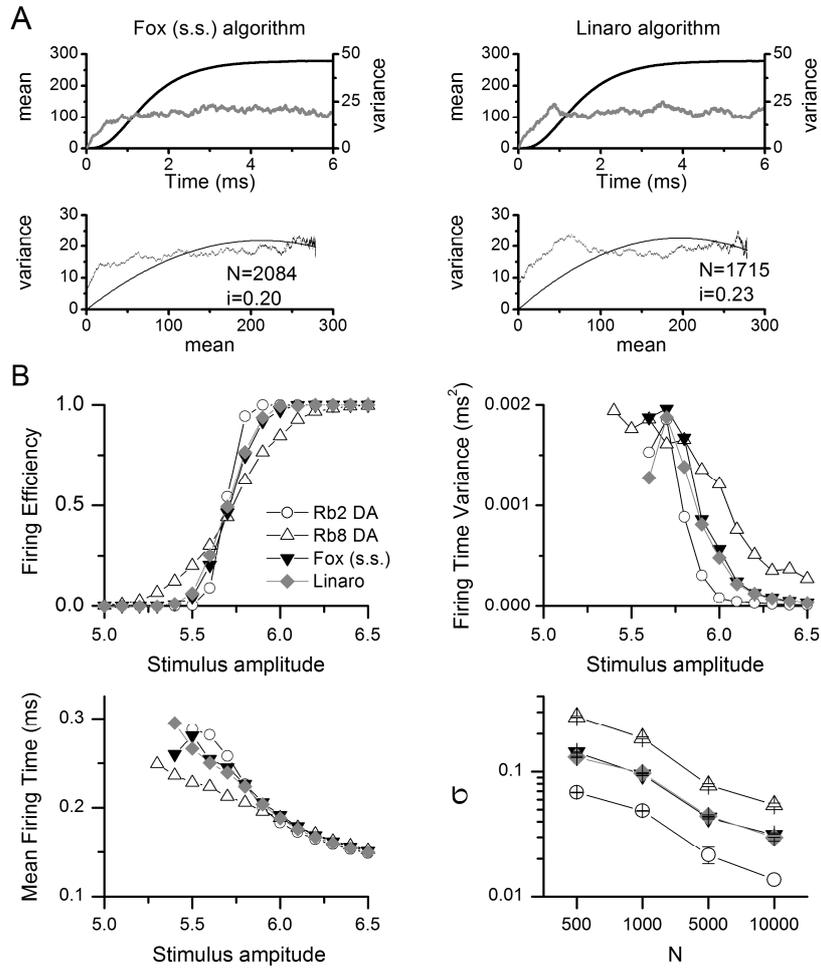

Figure 10

# SUPPLEMENTAL TEXT

Here we explain intuitively the method for building the stochastic differential equations (SDE) that will approximate any kinetic scheme for an ion channel, and derive the SDE for a sodium channel. We take as a working example the $m_1h_0$ state from the eight-state kinetic scheme for sodium channels (Supplemental Figure 1A)

*Deterministic terms*

For the deterministic (drift) terms of the stochastic differential equation, the six transitions that go from or come to the $m_1h_0$ state have to be considered. Each arrow represents a possible transition, its probability given by the product of the voltage-dependent kinetic constant times the value of the state that is at the beginning of the arrow. Terms given by arrows starting at $m_1h_0$ are negative (Supplemental Figure 1B) while the terms given by the arrows that end at $m_1h_0$ are positive (Supplemental Figure 1C). Thus, the six deterministic terms related to $m_1h_0$ are:

$$3\alpha_m m_0 h_0 - \beta_m m_1 h_0 - 2\alpha_m m_1 h_0 + 2\beta_m m_2 h_0 - \alpha_h m_1 h_0 + \beta_h m_1 h_1$$
$$= 3\alpha_m m_0 h_0 + 2\beta_m m_2 h_0 + \beta_h m_1 h_1 - \left(\beta_m + 2\alpha_m + \alpha_h\right) m_1 h$$

*Stochastic terms*

For the stochastic terms the transition arrows are to be considered in pairs. The pairs that are connected to $m_1h_0$ are (Supplemental Figure 1D):

$$3\alpha_m m_0 h_0 \,/\, \beta_m m_1 h_0$$
$$2\alpha_m m_1 h_0 \,/\, 2\beta_m m_2 h_0$$
$$\alpha_h m_1 h_0 \,/\, \beta_h m_1 h_1$$

Each pair of arrows originates a random term with zero mean and standard deviation equal to the square root of the sum of the transition probabilities, divided by the square root of $N_{Na}$, the number of sodium channels. In this case, all terms are considered positive. For the $m_1h_0$ state, the stochastic terms are

$$\xi_1 \sqrt{\frac{3\alpha_m m_0 h_0 + \beta_m m_1 h_0}{N_{Na}}} + \xi_2 \sqrt{\frac{2\alpha_m m_1 h_0 + 2\beta_m m_2 h_0}{N_{Na}}} + \xi_3 \sqrt{\frac{\alpha_h m_1 h_0 + \beta_h m_1 h_1}{N_{Na}}}$$

where $\xi_1$, $\xi_2$ and $\xi_3$ are independent Gaussian white noise terms with zero mean and unit variance. Note that each pair of arrows connects two (and only two) states. In the SDE for the second of such states, the stochastic term has to be repeated exactly (the same Gaussian term) but with the opposite sign. For instance, the transition pair $2\alpha_m m_1 h_0 / 2\beta_m m_2 h_0$ connects $m_1h_0$ and $m_2h_0$; therefore the SDE for the $m_2h_0$ state must contain the term

$$-\xi_2 \sqrt{\frac{2\alpha_m m_1 h_0 + 2\beta_m m_2 h_0}{N_{Na}}}$$

repeating the same Gaussian white noise term as in the $m_1h_0$ equation but with opposite sign. (being Gaussian terms with zero mean it doesn't matter which one goes positive; the key point is to have the term positive in one equation and negative in the other). Because of this, the full set of equations for the sodium channels has 20 stochastic terms but only 10 random variables; analogously the equations for a 5-state potassium channels have 8 stochastic terms with 4 random variables.

Following this procedure while keeping care of repeating stochastic term with opposite signs, the following set of equations for the sodium channel is obtained:

$$\frac{dm_0h_0}{dt} = \left(-3\alpha_m m_0 h_0 + \beta_m m_1 h_0 - \alpha_h m_0 h_0 + \beta_h m_0 h_1\right)$$
$$+\xi_1 \frac{1}{\sqrt{N_{Na}}}\sqrt{3\alpha_m m_0 h_0 + \beta_m m_1 h_0} + \xi_4 \frac{1}{\sqrt{N_{Na}}}\sqrt{\alpha_h m_0 h_0 + \beta_h m_0 h_1}$$

$$\frac{dm_1 h_0}{dt} = \left(3\alpha_m m_0 h_0 - \beta_m m_1 h_0 - 2\alpha_m m_1 h_0 + 2\beta_m m_2 h_0 - \alpha_h m_1 h_0 + \beta_h m_1 h_1\right)$$
$$-\xi_1 \frac{1}{\sqrt{N_{Na}}}\sqrt{3\alpha_m m_0 h_0 + \beta_m m_1 h_0} + \xi_2 \frac{1}{\sqrt{N_{Na}}}\sqrt{2\alpha_m m_1 h_0 + 2\beta_m m_2 h_0} + \xi_5 \frac{1}{\sqrt{N_{Na}}}\sqrt{\alpha_h m_1 h_0 + \beta_h m_1 h_1}$$

$$\frac{dm_2 h_0}{dt} = \left(2\alpha_m m_1 h_0 - 2\beta_m m_2 h_0 - \alpha_m m_2 h_0 + 3\beta_m m_3 h_0 - \alpha_h m_2 h_0 + \beta_h m_2 h_1\right)$$
$$-\xi_2 \frac{1}{\sqrt{N_{Na}}}\sqrt{2\alpha_m m_1 h_0 + 2\beta_m m_2 h_0} + \xi_3 \frac{1}{\sqrt{N_{Na}}}\sqrt{\alpha_m m_2 h_0 + 3\beta_m m_3 h_0} + \xi_6 \frac{1}{\sqrt{N_{Na}}}\sqrt{\alpha_h m_2 h_0 + \beta_h m_2 h_1}$$

$$\frac{dm_3 h_0}{dt} = \left(\alpha_m m_2 h_0 - 3\beta_m m_3 h_0 - \alpha_h m_3 h_0 + \beta_h m_3 h_1\right)$$
$$-\xi_3 \frac{1}{\sqrt{N_{Na}}}\sqrt{\alpha_m m_2 h_0 + 3\beta_m m_3 h_0} + \xi_7 \frac{1}{\sqrt{N_{Na}}}\sqrt{\alpha_h m_3 h_0 + \beta_h m_3 h_1}$$

$$\frac{dm_0 h_1}{dt} = \left(-3\alpha_m m_0 h_1 + \beta_m m_1 h_1 + \alpha_h m_0 h_0 - \beta_h m_0 h_1\right)$$
$$+\xi_8 \frac{1}{\sqrt{N_{Na}}}\sqrt{3\alpha_m m_0 h_1 + \beta_m m_1 h_1} - \xi_4 \frac{1}{\sqrt{N_{Na}}}\sqrt{\alpha_h m_0 h_0 + \beta_h m_0 h_1}$$

$$\frac{dm_1 h_1}{dt} = \left(3\alpha_m m_0 h_1 - \beta_m m_1 h_1 - 2\alpha_m m_1 h_1 + 2\beta_m m_2 h_1 + \alpha_h m_1 h_0 - \beta_h m_1 h_1\right)$$
$$-\xi_8 \frac{1}{\sqrt{N_{Na}}}\sqrt{3\alpha_m m_0 h_1 + \beta_m m_1 h_1} + \xi_9 \frac{1}{\sqrt{N_{Na}}}\sqrt{2\alpha_m m_1 h_1 + 2\beta_m m_2 h_1} - \xi_5 \frac{1}{\sqrt{N_{Na}}}\sqrt{\alpha_h m_1 h_0 + \beta_h m_1 h_1}$$

$$\frac{dm_2 h_1}{dt} = \left(2\alpha_m m_1 h_1 - 2\beta_m m_2 h_1 - \alpha_m m_2 h_1 + 3\beta_m m_3 h_1 + \alpha_h m_2 h_0 - \beta_h m_2 h_1\right)$$
$$-\xi_9 \frac{1}{\sqrt{N_{Na}}}\sqrt{2\alpha_m m_1 h_1 + 2\beta_m m_2 h_1} + \xi_{10} \frac{1}{\sqrt{N_{Na}}}\sqrt{\alpha_m m_2 h_1 + 3\beta_m m_3 h_1} - \xi_6 \frac{1}{\sqrt{N_{Na}}}\sqrt{\alpha_h m_2 h_0 + \beta_h m_2 h_1}$$

$$\frac{dm_3 h_1}{dt} = \left(\alpha_m m_2 h_1 - 3\beta_m m_3 h_1 + \alpha_h m_3 h_0 - \beta_h m_3 h_1\right)$$
$$-\xi_{10} \frac{1}{\sqrt{N_{Na}}}\sqrt{\alpha_m m_2 h_0 + 3\beta_m m_3 h_0} - \xi_7 \frac{1}{\sqrt{N_{Na}}}\sqrt{\alpha_h m_3 h_0 + \beta_h m_3 h_1}$$

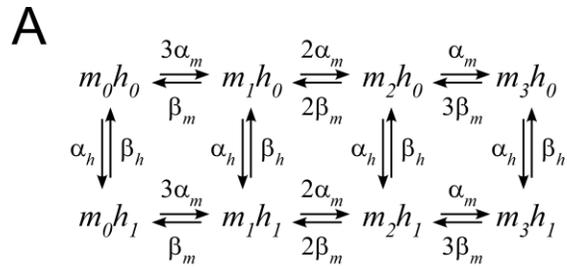

A

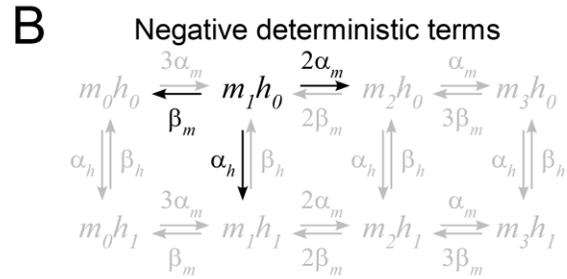

B Negative deterministic terms

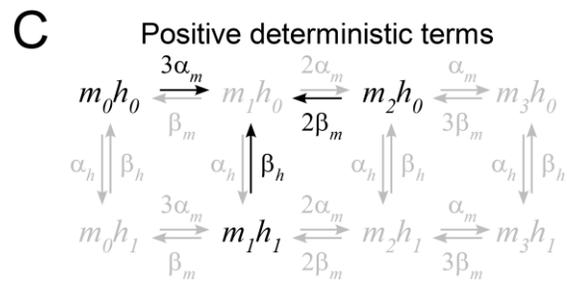

C Positive deterministic terms

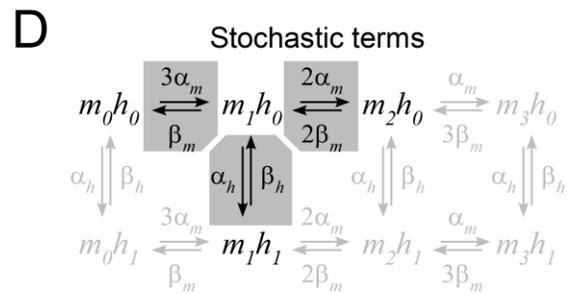

D Stochastic terms

Supplemental Figure 1